\documentclass{aa}
\usepackage{times}  
\input{epsf}

\begin{document}

\thesaurus{3(08.23.2;11.09.4;11.12.2;11.19.3;13.09.1;13.21.1)}

\title{ A Comprehensive Study of Intense Star Formation Bursts in Irregular
  and Compact Galaxies
\thanks{Based on observations from the {\em International Ultraviolet
Explorer} obtained at the ESA VILSPA observatory, on observations taken at
the Isaac Newton Telescope at the Spanish Observatorio del Roque de Los
Muchachos on La Palma island and on observations with the Nan\c cay
radiotelescope.}}

\author{J.~Miguel Mas-Hesse\inst{1}
\and Daniel Kunth\inst{2}}

\institute{Laboratorio de Astrof\'\i sica Espacial y F\'\i sica Fundamental
- INTA, POB 50727, E-28080 Madrid, Spain; e-mail: mm@laeff.esa.es
\and Institut d'Astrophysique de Paris, 98 bis Bd. Arago, F-75014 Paris, 
France; e-mail: kunth@iap.fr }

\offprints{J.~M. Mas-Hesse }

\date{Received December 1998 / Accepted March 1999}

\titlerunning{Intense star formation bursts in irregular and compact galaxies}
\authorrunning{Mas-Hesse \& Kunth}

\maketitle

\begin{abstract}

We have analyzed the properties of the star formation episodes taking place
in a sample of blue compact and irregular galaxies by comparing their
multiwavelength observational properties with the predictions of
evolutionary population synthesis models. This method has allowed us to
constrain the age, star formation regime (instantaneous or extended) and
Initial Mass Function (IMF) slope, as well as the shape and strength of the
interstellar extinction in these regions.  We  find that star
formation episodes are essentially short with  a mean age of 3.5
Myrs. Some galaxies may be undergoing their first global episode of star
formation while for the rest of the sample older stars contribute to at
most half the optical emission.  The Wolf-Rayet star population (WR) is well
reproduced by the models and provides the strongest argument in favor of a
short duration of the star formation episode.  Supernova rates are
relatively large. The accumulation of supernova explosions within few Myr
has contributed to a quick metal enrichment of the ISM and to its
disruption by the release of huge amounts of mechanical energy. $V-K$ colors
agree well with the prediction that red supergiant stars are rare in low
metallicity regions. A general agreement is found between the predicted and
observed far infrared emissions suggesting that the fraction of hidden
stars contributing to the ionisation is minimum, except in some specific
objects. A saillant result of this study is that the IMF slope appear to be
very universal, on average very close to that of the solar neighborhood and
with no dependence on the metallicity, contrary to previous claims. We have
also found no dependence whatsoever between the shape of the extinction law
and the metallicity. It is likely that the strong radiation associated to
the bursts destroys the dust component responsible for the 2175 \AA\ bump.
 Finally we confirm that extinction affecting the
stellar continuum is in some cases significantly weaker than that derived
from the Balmer emission lines. Such a discrepancy can lead to
underestimations in the value of  the {H$\beta$} equivalent width by a factor as
 large as 2, leading
to an overestimation of the age of the burst. Similarly, 
the Wolf-Rayet bump to the  {H$\beta$} luminosities ratio
can also be affected by this differential reddening leading to an
overestimation of the WR star population. As bursts get older they appear
dustier, possibly as a result of dust ejection during the evolution of
their most massive stars. Finally, we have found a serious general
discrepancy between the predicted and the measured radio
luminosities. While part of this discrepancy might be attributed to
aperture mismatching in some cases, it points to the presence of additional
radio sources not included in present evolutionary models. 

\keywords{Stars: Wolf-Rayet; Galaxies: ISM; Galaxies: luminosity function,
  mass function; Galaxies: starburst; Infrared: galaxies; Ultraviolet:
  galaxies}

\end{abstract}

\section{Introduction}

With the launch of the HST we are now witnessing a new era that allows to
analyze in detail nearby objects such as Galactic \ion{H}{ii} regions or 30 Dor in
the LMC. Such studies permit to resolve and study individual stars in
massive star clusters and obtain UV spectra of hot stars with metallicities
that are not solar. These studies combined with follow up observations from
10-m class telescopes will bring an enormous knowledge about massive star
formation and evolution processes and in particular direct informations
about the IMF slope, as well as its lower and upper mass limits and
velocity dispersion of individual star clusters.  However, as one tackles
with objects at larger distances, stars will remain unresolved and so any
argument will still be based on global properties. In particular in the
quest for primeval galaxies at very large redshift one must be prepared to
identify them and a thorough knowledge of the properties of nearby
starbursts is indeed mandatory. For this reason studies of massive star
formation in nearby galaxies have received a great attention in the past
years. In the UV the seminal paper of Meier \& Terlevich (1981) has
pointed out the possible resemblance of a primeval galaxy with some local
blue compact galaxies that might be forming stars for the first
time. Kinney et al. (1996) compiled a set of templates from the UV to the
near infrared for galaxies of different kinds, including starbursts,
spirals and ellipticals, aiming to provide a way to properly apply
K-corrections for high redshift objects.  Very recently, Heckman et
al. (1998) have presented a systematic study of the UV properties of a
sample of 45 starburst and related galaxies, proposing to use them as a
``training set'' to better understand the properties of starbursts at high
redshift. In parallel, efforts have been put in building various models
using population synthesis techniques to reproduce all the observed
electromagnetic spectrum from the UV up to X-rays (see the proceedings of
the 1995 IAP (Kunth et al. 1996), 1995 Crete (Leitherer et al. 1996) and
1998 Puerto Vallarta (van der Hucht et al. 1998) conferences for a
comparative compilation of synthesis models and observational results).

With these ideas in mind we started some years ago the detailed study
of massive star formation episodes in different environments. The
predictions of evolutionary populations synthesis models at various
metallicities were presented and discussed in Arnault, Kunth \& Schild 
(1989),
Mas-Hesse \& Kunth (1991a) and Cervi\~no \& Mas-Hesse (1994). We now
present a comprehensive observational study of massive star--formation
episodes taking place in star--forming galaxies and Giant \ion{H}{ii} regions
in nearby spiral galaxies. Preliminary partial results have been discussed in
Mas-Hesse (1990) and Mas-Hesse \& Kunth (1991b, 1996a,b).  
Our analysis is based on the integrated emission from the whole
massive star clusters. The sample comprises a set of galaxies with
different morphological types, luminosities and hosting environments. Some
of them are close enough to have been recently studied with the HST,
allowing comparisons to be made, while others are far more distant and
remain suited for integrated studies like the present one or for
statistical investigations.  Our goal is to characterize what are the universal
properties of starbursting episodes: Initial Mass Function (IMF), star
formation regime (instantaneous or extended), evolutionary state, effects
of dust particles, effects on the interstellar medium (ISM), etc... To
achieve this, IUE UV data have been complemented with optical
spectroscopical observations, with special care to match both apertures.
The data set has been completed with IR, far infrared (FIR) and radio
data. We will show that the comparison of multiwavelength observational
data with the predictions of synthesis models is a powerful tool to
constrain the properties of these star--forming episodes.

The observational data (optical, ultraviolet, FIR and radio) are described
in Sect. 2; the predictions of our evolutionary models are presented in
Sect. 3; the model fitting techniques require some subtle handling that
critically depends on the quality and completeness of each available data
set as exposed in Sect. 4; the results are given in Sect. 5, together with
a complete discussion of the processes that govern fundamental starburst
properties; in Sect.~6 we discuss in more detail the properties of some
individual galaxies; our conclusions are finally summarized in Sect. 7.

\begin{figure}
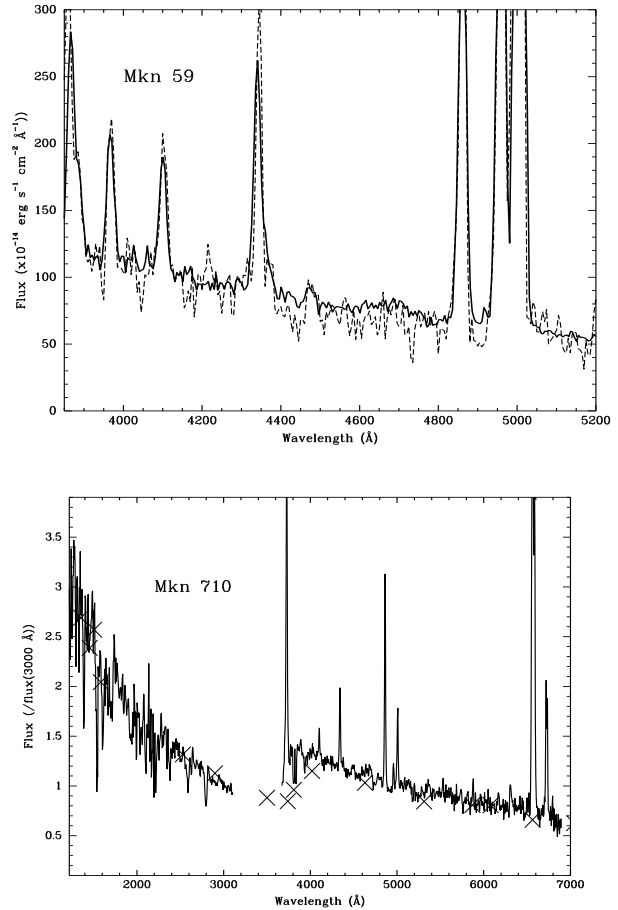
 
\begin{center}\mbox{\epsfxsize=8cm \epsfbox{ms8377.f1a}}\end{center}
\begin{center}\mbox{\epsfxsize=8cm \epsfbox{ms8377.f1b}}\end{center}
\caption[]{
a) Mrk~59 IPCS spectrum (solid line) over the corresponding CCD 
spectrum, both taken with 10\arcsec\ broad slits. Note the good photometrical 
agreement between both spectra, obtained during different runs and with 
different detectors. 
b) UV -- optical spectrum (crosses) of Mrk~710 from Storchi-Bergmann et al. 
(1995) over the IUE and optical spectra obtained by us (solid line). The 
agreement of the optical spectra is rather good, especially above 4000~\AA\
(the continuum of our optical spectrum is quite noisy below 3900~\AA).   
}
\label{comp}
\end{figure}

\begin{table*}
\caption[]{Adopted properties of the selected star--forming regions. 
$H_0 = 75$ km s$^{-1}$ Mpc$^{-1}$ has been assumed. Coordinates have been
taken from the NASA Extragalactic Database (NED). Galactic {$E(B-V)$} values
have been taken from Burstein \& Heiles (1984).}
\begin{flushleft}
\begin{tabular}{lcccccc}
\hline
Name  & R.A.  & Dec.  & Distance & Morphology & $E(B-V)_{Galactic}$ & Other names\\
      &   (1950.0)  &  & (Mpc) &  &                 &   \\
\hline
\object{NGC~588}  & 01 29 58.5 &  +30 23 51 & 0.82$^1$  & G\ion{H}{ii} & 0.045 & In M33  \\
\object{NGC~595}  & 01 30 44.7 &  +30 26 08 & 0.82$^1$  & G\ion{H}{ii} & 0.045 & In M33  \\
\object{NGC~2363} & 07 23 23.7 &  +69 17 33 & 3.0$^1$   & IG   & 0.042 & In \object{NGC~2366}\\
\object{IZw~18}   & 09 30 30.2 &  +55 27 49 & 10.7$^2$  & BCDG & 0.005 & \object{Mrk~116}  \\
\object{Mrk~710}  & 09 52 10.2 &  +09 30 32 & 17.9$^3$  & SG   & 0.010 & \object{NGC~3049} \\
\object{Tol~3}    & 10 04 18.2 &  -29 41 29 & 11.6$^4$  & BCDG & 0.062 & \object{NGC~3125} \\
\object{NGC~3256} & 10 25 43.4 &  -43 38 48 & 50.0$^5$  & Merger & 0.14 & \object{VV 065} \\ 
\object{Haro~2}   & 10 29 22.7 &  +54 39 31 & 20.2$^6$  & BCDG & 0.0   & \object{Mrk~33}  \\
\object{Mrk~36}   & 11 02 15.6 &  +29 24 28 & 8.0$^7$   & BCDG & 0.0   & \object{Haro~4}  \\
\object{NGC~4214} & 12 13 08.0 &  +36 36 22 & 6.4$^8$   & IG   & 0.0   &         \\
\object{IZw~36}   & 12 23 50.5 &  +48 46 13 & 4.6$^9$   & BCDG & 0.0   & \object{Mrk~209} \\
\object{NGC~4670} & 12 42 49.9 &  +27 23 55 & 12.1$^6$  & BCDG & 0.005 & \object{Haro 9}  \\
\object{Mrk~59}   & 12 56 38.5 &  +35 06 56 & 11.7$^6$  & IG   & 0.0   & \object{NGC~4861} \\
\object{NGC~5253} & 13 37 05.0 &  -31 23 30 & 4.1$^{10}$   & IG   & 0.047 & \object{Haro~10}  \\
\object{NGC~5471} & 14 02 42.7 &  +54 38 08 & 6.3$^{11}$   & G\ion{H}{ii} & 0.0   & In M101  \\
\object{IIZw~70}  & 14 48 55.1 &  +35 46 37 & 17.6$^7$  & BCDG & 0.0   & \object{Mrk~829}  \\
\object{IC~4662}  & 17 42 12.0 &  -64 37 18 & 4.0$^{12}$   & IG   & 0.065 & \object{He~2-269} \\ 
\hline
\end{tabular} 
\end{flushleft} 
\label{sample} 
Notes- G\ion{H}{ii}: giant \ion{H}{ii} region; IG: irregular galaxy; BCDG: blue compact 
dwarf galaxy; SG: starburst galaxy.  \\
Distance references- $^1$Sandage \& Tamman (1974a), $^2$Kinman \& Davidson (1981),
$^3$Mazzarella \& Balzano (1986), $^4$Joseph \& Wright (1985), $^5$Kunth \& S\`evre 
(1986),  $^6$Huchra et al. (1983), $^7$French (1980), $^8$Sandage \& Tamman (1974b), 
$^9$Viallefond \& Thuan (1983), $^{10}$Saha et al. (1995), $^{11}$Aaronson \&
Mould (1983), $^{12}$Alloin \& Sareyan (1974). 
\end{table*}

\section{Observational data}

        We selected galaxies with \ion{H}{ii}-region-like emission-line spectra and
with a strong UV continuum, so that they could be observed with IUE.  12
objects were taken from the IUE catalogue of extragalactic \ion{H}{ii} regions from
Rosa et al.  (1984), rejecting those observations with low signal-to-noise
or with technical problems (saturation, read-out errors...).  Haro~2,
IZw~36 and NGC~4670 were included for comparison with previous results by
other investigators.  We also included Mrk~710 and NGC~3256.  Mrk~710 is a
starburst galaxy with an important population of Wolf-Rayet stars (Kunth \&
Schild 1986), which is interesting to test the age calibration based on other
parameters.  NGC~3256 is an ultraluminous IRAS galaxy ({$L_{FIR}$} ~$\approx ~2~
10^{45}$ erg s$^{-1}$), probably a ``merger'' (Graham et al. 1984), and
was selected to analyze if its enormous FIR luminosity was due to
 a very intense starburst triggered by the collision.

We list in Table~\ref{sample} the name, coordinates, distance and 
morphological type of the 17 objects included in our sample. Distances have 
been taken from the literature, as indicated in the table. When needed, they 
were corrected so that all of them are consistent with $H_0 = 75$ km
s$^{-1}$ Mpc$^{-1}$. The Galactic {$E(B-V)$} color
excess derived by Burstein \& Heiles (1984) for each galaxy is also listed
in Table~\ref{sample}. Seven are blue compact dwarf galaxies, while other 5 are
irregular galaxies. The sample is completed with three giant \ion{H}{ii} regions, a
starburst nucleus and an ultraluminous IRAS galaxy for comparison.

We have compiled UV, optical, far infrared and radio data for the majority
of the objects in the sample.  We have combined data taken from the
literature and from astronomical databases (IUE and IRAS) with new
observations, in order to cover a spectral range as wide as possible.

\begin{table*}
\caption[]{Log of optical observations. An WG3650 filter was used with 
the IPCS to minimize second order contamination at H$\alpha$. }
\begin{flushleft}
\begin{tabular}{lcccccc}
\hline
Object   &     Date   &  Detector    & Slit & Pos. angle & Exp. time & $\lambda$ range \\
         &            &              &   ($\arcsec$)  & (deg.) & (s) &     \AA \\
\hline
IZw~18   &  23-01-88  & INT+IPCS      &  10  & 135  & 1500 &  3500--7000 \\
Mrk~710  &  22-01-88  & INT+IPCS      &  10  & 30   & 1800 &  3500--7000 \\
Tol~3    &  23-02-80  & LC+RETI       &  1x2 &  --  & 4800 &  3500--6900 \\
Haro~2   &  22-01-88  & INT+IPCS      &  10  & 330  & 1200 &  3500--7000 \\
         &  22-01-88  & INT+IPCS      &  10  & 330  & 1300 &  3500--7000 \\
         &  11-05-88  & INT+CCD       &  2   & 330  & 4500 &  3600--5300 \\
         &  12-05-88  & INT+CCD       &  2   & 330  & 2500 &  3600--5300 \\
         &  12-05-88  & INT+CCD       &  10  & 330  & 600  &  3600--5300 \\
Mrk~36   &  23-01-88  & INT+IPCS      &  10  & 90   & 660  &  3500--7000 \\
NGC~4214 &  22-01-88  & INT+IPCS      &  10  & 120  & 600  &  3500--7000 \\
         &  11-05-88  & INT+CCD       &  2   & 285  & 3000 &  3600--5300 \\
         &  12-05-88  & INT+CCD       &  10  & 285  & 300  &  3600--5300 \\
         &  12-05-88  & INT+CCD       &  2   & 285  & 1500 &  4250--5950 \\
IZw~36   &  23-01-88  & INT+IPCS      &  10  & 90   & 2220 &  3500--7000 \\
NGC~4670 &  22-01-88  & INT+IPCS      &  10  & 100  & 1500 &  3500--7000 \\
         &  11-05-88  & INT+CCD       &  2   & 265  & 2800 &  3600--5300 \\
         &  12-05-88  & INT+CCD       & 10   & 265  & 360  &  3600--5300 \\
Mrk~59   &  23-01-88  & INT+IPCS      & 10   & 10   & 1200 &  3500--7000 \\
         &  12-05-88  & INT+CCD       & 10   & 10   & 300  &  3600--5300 \\
         &  12-05-88  & INT+CCD       & 2    & 10   & 2000 &  3600--5300 \\
NGC~5253 &  31-01-84  & LS+IDS        & 4x4  & --   & 600  &  3500--5900 \\
IIZw~70  &  11-05-88  & INT+CCD       & 2    & 225  & 3700 &  3600--5300 \\
         &  11-05-88  & INT+CCD       & 10   & 225  & 1200 &  3600--5300 \\  
\hline
\end{tabular} 
\end{flushleft} 
\label{log1} 
Notes- INT: Isaac Newton Telescope; LC: Las Campanas 2.5m telescope; LS: La Silla 
3.6m telescope.  
\end{table*}

\subsection{ Optical} 

The main goals of the optical observations were essentially:

\begin{itemize}

\item  To get the continuum spectral distribution in the 3600--\-7000 \AA\
range over the same aperture as IUE ($10\arcsec\times 20\arcsec$). Aperture
mismatch would have made impossible to compare directly the UV and optical
continua. Moreover, by using such a large aperture we generally covered the
major part of the ionized regions, which are much more extended than the
stellar clusters, allowing us to compare the spatially integrated emission
of the Balmer lines with model predictions. In some objects Balmer lines
fluxes can be underestimated by more than a factor 10 if obtained through
narrow slits only $1\arcsec$ wide.

\item To observe with good signal-to-noise ratio the continuum around
4650~\AA , looking for broad emission lines which could be the signature of
Wolf-Rayet stars.

\end{itemize}

Nine objects of the sample were observed in January and May~1988 with the
2.5~m Isaac Newton telescope at La Palma Observatory, using the
Intermediate Dispersion Spectrograph on the Cassegrain focus.  We summarize
in Table~\ref{log1} the log of optical observations.

The January 1988 observations were performed with an Image Photon Counting
System (IPCS) and a 138.5~\AA /mm grating, covering the spectral range
3600--7000~\AA\ with 2.0 \AA\ per 15~$\mu$m pixel.  We used a WG360 filter
to cut the incident radiation below 3600~\AA , avoiding so the
contamination of the red part of the spectrum by second-order light.  We
took care to limit the count rate below 0.2 cts/s by using
the corresponding neutral density filters, since the IPCS looses linearity
 at higher frequencies.  The slit width was set to
$\approx$~10$\arcsec$ in order to match the IUE aperture and cover most of the
 star-forming regions.

The data from May 1988 were obtained with a blue-coated GEC P8603 CCD 
detector and the same instrumental configuration, observing
the range 3600--5300 \AA\ with 3.0 \AA\ per 22~$\mu$m pixel.  With a slit 
width of 1.8$\arcsec$ and allowing for relatively long
integration times we got well exposed continua in the region of the 
Wolf-Rayet features. We also took some spectra with a 10$\arcsec$ slit
to check the photometric calibration of the IPCS spectra taken in January.
We finally completed the data sample with narrow slit data for Tol~3 and
NGC~5253.  

The data were reduced using the European Southern Observatory IHAP package
using standard procedures which included flat field corrections, bias
subtraction (CCD), geometrical distortion correction (IPCS), wavelength
calibration, sky subtraction and conversion to absolute fluxes.  Three
standard stars were observed each night through a 5$\arcsec$ slit.  The
instrumental response curves derived were very stable, with differences
smaller than 10\%, so that we used a mean curve for each observing run.  We
compare in Fig.~\ref{comp}a the optical IPCS and CCD spectra of Mrk~59,
both obtained through the same broad slit during different runs.  The
discrepancies between the IPCS and CCD spectra taken with the same slit
were found in any case to be smaller than 10\%, precision that can be
assigned to our photometrical calibration.  We performed a further check of
our calibration by comparing our UV - optical spectra with those obtained
by Storchi-Bergmann et al. (1995) and McQuade et al. (1995), also obtained
through a broad slit consistent with the IUE aperture. In Fig.~\ref{comp}b
we plot the data corresponding to Mrk~710. It can be seen that both
calibrations are consistent, and that the optical spectra are similar (the
UV spectra used by these authors are the same IUE ones obtained by us). The
agreement is quite good above 4000~\AA, while at shorter wavelengths the
low sensitivity of the IPCS yielded quite noisy continua.  The spectra were
corrected from Galactic extinction taking the color excesses from Burstein
\& Heiles (1984) and using the extinction law as parameterized by Seaton
(1979).  Spectra were finally centered to nominal wavelengths, considering
that the observed wavelength of a line was affected not only by
cosmological redshift, but also by the location of the object in the rather
wide 10$\arcsec$ slit.

\begin{table*}
\caption[]{Optical observed and derived properties. We list here the 
full emission lines intensities for the objects we did observe. 
Otherwise, derived data have been taken from the literature. {$W(H\beta)$} and
I(WR)/{H$\beta$} values are
given as measured on the spectra, without any correction.  }
\begin{flushleft}
\begin{tabular}{lccccccccc}
\hline
       &     NGC~588     &  NGC~595 & NGC~2363 & IZw~18 & Mrk710 & Tol~3 & NGC~3256 & Haro~2 &
 Mrk~36 \\
\hline
I([\ion{O}{ii}]$\lambda$3727)   & --  & -- &  -- &  0.49 & 1.76 & 0.87 & -- & 3.70 & 1.82 \\
I(H$\delta$)   & -- &  -- &  -- &  0.32 & 0.09 & 0.19 & -- & 0.12 & 0.19 \\
I(H$\gamma$)   & -- &  -- &  -- &  0.48 & 0.36 & 0.39 & -- & 0.37 & 0.39  \\
I([\ion{O}{iii}$\lambda$4363)   & -- & -- & -- & 0.04 & -- & 0.04 & -- & -- & 0.09  \\
I(H$\beta$)   &  -- &  -- &  -- & 1.00 & 1.00 & 1.00 & -- & 1.00 & 1.00 \\
I([\ion{O}{iii}]$\lambda$4959)   &  -- & -- &  -- & 0.60 & 0.13 & 1.95 & -- & 0.62 & 1.54 \\
I([\ion{O}{iii}]$\lambda$5007)   &  -- & -- &  -- & 2.06 & 0.40 & 5.76 & -- & 2.14 & 5.05 \\
I(H$\alpha$)   &  -- &  -- &  -- &  2.97 & 4.25 &  2.41 & -- & 3.79 & 2.84 \\
I(WR $\lambda$4650)    & -- &  -- & -- & 0.0 & 0.24 & 0.09 & -- & 0.15 & 0.0  \\
{$E(B-V)$}     &  0.10$^a$     &  0.33$^a$ &  0.16$^c$ &  0.05 & 0.29 & 0.17 & 0.85$^d$ & 0.22 & 0.0\\
{$L(H\beta)$} (obs.)     &  $6.5\,10^{37}$$^b$  & $1.4\,10^{38}$$^b$ & $9.2\,10^{38}$$^c$ & $1.3\,10^{39}$ &
 $6.7\,10^{39}$ & $1.1\,10^{39}$ & $1.1\,10^{41}$$^d$ & $2.0\,10^{40}$ & $7.3\,10^{38}$\\
{$L(H\beta)$} (dered.)&  $1.2\,10^{38}$  & $4.5\,10^{38}$ & $1.6\,10^{39}$ & $1.5\,10^{39}$ & 
$1.8\,10^{40}$ & $2.0\,10^{39}$ & $1.9\,10^{42}$ & $4.2\,10^{40}$ & $7.3\,10^{38}$ \\
(erg/s) &&&&&& $(1.6\,10^{40}$$^d$)\\
{$W(H\beta)$} (\AA) &  97$^a$ &  $\approx$100$^a$ &  300$^c$ &  135 & 35 & 98 & 20$^d$ & 30 & 110 \\
       &     &               &      &      &    & (50$^d$) & & &\\
$O/H$  & 8.3 & 8.4 & 7.9 & 7.2 & 8.9 & 8.2 & 8.9 & 8.4 & 7.8 \\
{$T_{eff}$}   &  36600 &  35000 &  42700 &  35000 & 39800 & 38200 & 37000 & 35400 & 40000 \\
$L_{3000}$ &  $9.66\,10^{35}$ &  $9.50\,10^{35}$ &  $1.94\,10^{37}$ &  $3.70\,10^{37}$ &
$1.97\,10^{38}$ & $2.22\,10^{38}$ & $1.20\,10^{40}$$^e$ & $6.55\,10^{38}$ & $6.81\,10^{37}$$^e$ \\
(erg s$^{-1}$ \AA$^{-1}$)&&&&&&\\
\hline
\end{tabular} 
\end{flushleft} 
\label{opti} 
\end{table*}

\addtocounter{table}{-1}

\begin{table*}
\caption[]{{\it Continuation.}}
\begin{flushleft}
\begin{tabular}{lcccccccc}
\hline
       & NGC~4214  &  IZw~36 & NGC~4670 & Mrk~59 & NGC~5253 & NGC~5471 & IIZw~70 & IC~4662 \\
\hline
I([\ion{O}{ii}]$\lambda$3727)   & 4.44 & 0.86 & 3.40 & 1.68 & 1.61 & -- & 2.07 & -- \\
I(H$\delta$)   & 0.11 & 0.25 & -- & 0.24 & 0.20 & -- & 0.14 & -- \\
I(H$\gamma$)   & 0.38 & 0.62 & 0.40 & 0.40 & 0.41 & -- & 0.42 & -- \\
I([\ion{O}{iii}]$\lambda$4363)   & -- & -- & -- & -- & 0.04 & -- & -- & -- \\
I(H$\beta$)   &  1.00 & 1.00 & 1.00 & 1.00 & 1.00 & -- & 1.00 & -- \\
I([\ion{O}{iii}]$\lambda$4959)   &  1.02 & 1.99 & 1.11 & 2.03 & 1.75 & -- & 1.29 & -- \\  
I([\ion{O}{iii}]$\lambda$5007)   &  3.21 & 6.82 & 3.29 & 6.82 & 5.62 & -- & 3.69 & -- \\
I(H$\alpha$)   &  4.01 & 3.03 & 4.29 & 2.86 & -- & -- & -- & -- \\
I(WR $\lambda$4650)    & 0.25 & -- & $\le$0.1 & 0.08 & 0.03 & 0.02 & 0.0 & 0.005 \\   
{$E(B-V)$}     &  0.28 & 0.0 & 0.32 & 0.0 & 0.19 & 0.15$^f$ & 0.17 &  0.09$^i$ \\
{$L(H\beta)$} (obs.)     &  $4.9\,10^{39}$ & $8.6\,10^{38}$ & $7.0\,10^{39}$ & $1.4\,10^{40}$ &
$1.8\,10^{38}$& $1.2\,10^{39}$$^g$ & $7.8\,10^{39}$ & $1.6\,10^{39}$$^h$ \\
{$L(H\beta)$} (dered.) &  $1.3\,10^{40}$ & $8.6\,10^{38}$ & $2.1\,10^{40}$ & $1.4\,10^{40}$ &
$3.4\,10^{38}$ & $2.0\,10^{39}$ & $1.4\,10^{40}$ & $2.2\,10^{39}$ \\
(erg/s)&&&&&$(6.9\,10^{39}$$^d$)&&&\\
{$W(H\beta)$}  (\AA) & 49  & 300 & 27 & 135 & 161 & 129$^g$ & 49 & 65$^h$ \\
       &     &     &    &     & (80$^d$)&  & & \\
$O/H$  & 8.4 & 7.8 & 8.4 & 8.0 & 8.2 & 8.0 & 8.0 & 8.3 \\
{$T_{eff}$}   &  37000 & 40000 & 37000 & 40000 & 37000 & 39000 & 36000 &  38000 \\  
$L_{3000}$ &  $1.33\,10^{38}$ & $2.73\,10^{37}$$^e$ & $3.36\,10^{38}$ & $2.83\,10^{38}$ & 
$1.12\,10^{38}$ & $6.84\,10^{37}$ & $2.74\,10^{38}$ & $6.37\,10^{37}$ \\
(erg s$^{-1}$ \AA$^{-1}$)&&&&&&\\
\hline
\end{tabular} 
\end{flushleft} 
\label{opti2} 
Notes and references: $^a$~V\'{\i}lchez et al. (1988); $^b$~Melnick (1979), the slit comprises the 
whole \ion{H}{ii} region;  $^c$ Peimbert et al. (1986), 4$\arcsec$x12$\arcsec$ slit; $^d$ Storchi-Bergmann
et al. (1995), $10\arcsec\times 20\arcsec$ slit; $^e$ continuum luminosity measured at 1900~\AA; $^f$ Rosa 
\& Benvenuti (1994); $^g$ Torres-Peimbert et al. (1989); $^h$ Alloin (1974), 
 47$\arcsec$ circular aperture ; $^i$ Heydari-Malayeri et al. (1990). 
\end{table*}

We list in Table~\ref{opti} the intensity of several lines measured on the
10$\arcsec$ slit spectra. We also list the intensity of the Wolf-Rayet bump
measured on the narrow slit spectra.  This bump is a blend of broad stellar
atmospheric emission lines: \ion{He}{ii}~$\lambda$4686, \ion{N}{v}~$\lambda$4603,
$\lambda$4619, \ion{N}{iii}~$\lambda$4634, $\lambda$4640, $\lambda$4642,
\ion{C}{iii}~$\lambda$4650 and \ion{C}{iv}~$\lambda$4658, where the nitrogen emission lines
are attributed to stars of the WN sequence and the carbon emission to WC
sequence stars, respectively.  The WR bump intensity has been obtained
interpolating the continuum between 4600 and 4700~\AA\ and integrating all
the excess emission above.  The contribution of obvious narrow nebular
lines has been subtracted by fitting a gaussian.  In any case, of equal
importance is the very detection or non-detection of the bump, indicative
of the presence of Wolf-rayet stars in the cluster.

A first estimation of the extinction was derived by comparing the observed
ratios between some Balmer lines ({H$\alpha$} ~to {H$\delta$} ) and  theoretical values
computed by Osterbrock (1989).  Significant discrepancies in the derived
color excesses {$E(B-V)$} were found depending on the pair of lines selected.
Similar discrepancies have already been found by other authors in the past
(Kunth \& Sargent, 1983; Campbell et al. 1986) and are usually attributed to
stellar absorption lines contamination, with equivalent widths in the range
1--2~\AA .  For consistency with previous works we have fixed the {H$\beta$}
~absorption equivalent width to 1~\AA , and have estimated which correction
were needed in the {H$\gamma$} ~and {H$\delta$} ~lines to derive an homogeneous color
excess.  We have found that assuming absorption equivalent widths in the
range 1--4~\AA , with a mean value of 1.8~\AA , the same {$E(B-V)$} is derived
independently of the pair of lines used.  This value of {$E(B-V)$} has been
listed in Table~\ref{opti}. The correction to the emission flux
calculated in this way amounts in average to 2\% for {H$\beta$}, 9\% for {H$\gamma$} ~and
50\% for {H$\delta$}.

Oxygen abundances had already been published by other authors in the past
but have been recalculated for those galaxies we observed in the optical.
The measured oxygen abundances are listed in Table~\ref{opti}. We have
completed it with values taken from the literature.  Effective
temperatures were estimated following the method described in Cervi\~no \&
Mas-Hesse (1994), based on the presumed  temperature dependence of the
 oxygen over {H$\beta$} lines ratios.

The optical spectroscopy was complemented with  data gi\-ven by
Storchi-Bergmann et al. (1995), which 
were also taken thro\-ugh broad slits consistent with the IUE one.  

\begin{table}
\caption[]{Measured \ion{Si}{iv} and \ion{C}{iv} absorption lines equivalent widths.}
\begin{flushleft}
\begin{tabular}{lccc}
\hline
Object   & $W(\ion{Si}{iv})$ & $W(\ion{C}{iv})$ & $W(\ion{Si}{iv})/W(\ion{C}{iv})$\\
         &    \AA    &  \AA     &                 \\
\hline
NGC~588  &  0.7$\pm$0.4 & 5.0$\pm$1.0 &  0.14$\pm$0.14\\
NGC~595  &  3.1$\pm$0.8 & 5.7$\pm$0.4 &  0.54$\pm$0.10\\
NGC~2363 &    em     & em       &  -- \\
IZw~18   &    --     &   --     &  --  \\
Mrk~710  &  6.0$\pm$1.0 & 8.5$\pm$0.6 &  0.71$\pm$0.10\\
Tol~3    &  1.8$\pm$0.3 & 4.0$\pm$0.5 &  0.45$\pm$0.12\\
NGC~3256 &  7.5$\pm$1.0 & 7.5$\pm$2.0 &  1.0$\pm$0.3 \\
Haro~2   &  5.4$\pm$0.7 & 5.5$\pm$0.7 &  1.0$\pm$0.2\\
Mrk~36   &  0.6$\pm$0.1 & 3.0$\pm$0.5 &  0.20$\pm$0.15\\
NGC~4214 &  2.8$\pm$0.3 & 4.2$\pm$0.4 &  0.67$\pm$0.12\\
IZw~36   &  em       & em       & -- \\
NGC~4670 &  4.1$\pm$0.4 & 5.0$\pm$0.6 &  0.82$\pm$0.15\\
Mrk~59   &  2.0$\pm$0.6 & 2.2$\pm$0.4 &  0.91$\pm$0.30\\
NGC~5253 &  2.2$\pm$0.4 & 3.7$\pm$0.5 &  0.59$\pm$0.17\\
NGC~5471 &  1.0$\pm$0.2 & 2.0$\pm$0.2 &  0.50$\pm$0.15\\
IIZw~70  &  1.2$\pm$0.2 & 4.0$\pm$0.6 &  0.30$\pm$0.10\\
IC~4662  &  2.9$\pm$0.5 & 2.9$\pm$0.5 &  1.0$\pm$0.3\\ 
\hline
\end{tabular} 
\end{flushleft} 
\label{iue2} 
Notes- em: absorption profiles contaminated by nebular emission. 
\end{table}

\subsection{ Ultraviolet} 

We have compiled ultraviolet spectra taken with the International
Ultraviolet Explorer (IUE) for all the galaxies in our sample.  General
information and references about IUE can be found in Kondo (1988).  
All these spectra were taken at low resolution
($\approx$6~\AA ) through an oval entrance aperture of approximately
$10\arcsec\times 20\arcsec$.  The whole 1100--3200~\AA\ spectral range
is available for the majority of the galaxies except NGC~3256, Mrk~36 and
IZw~36, for which only the short wavelength range has been observed.

\begin{table*}
\caption[]{FIR and radio luminosities. IZw~18 and IZw~36 were not
  detected by IRAS. The upper limit indicated corresponds to the
  maximum flux expected according to the sensitivity of the IRAS detectors. }
\begin{flushleft}
\begin{tabular}{lrrrcrcr}
\hline
Object & IRAS source & $f_{60}$ & $f_{100}$ & $T_{dust}$ & {$L_{FIR}$} & {$L_{radio}$} & {$\alpha_{radio}$} \\
       &      &    (Jy)  &   (Jy)    &  (K) &  (erg/s) &  (Jy kpc$^2$) & (6-20 cm) \\
\hline
NGC~588   &  01299+3023 & 1.0   & $\le$5.4  &   --  & $\le1.2\,10^{40}$  & --  & -- \\      
NGC~595   & F01307+3025 & 11.0  & $\le$35.1 &   --  & $\le8.3\,10^{40}$  & $6.3\,10^4$$^a$ & $\approx$0  \\
NGC~2363  &  07233+6917 & 3.3   & 4.6&  42   & $2.7\,10^{41}$  & $1.3\,10^6$$^b$  & $\approx$0 \\
IZw~18    &  --         &  --   & --  &   --  & $<6.5\,10^{41}$   & $2.3\,10^6$$^b$  & -0.12  \\
Mrk~710   &  09521+0930 & 2.7   & 4.3 &  40 & $7.4\,10^{42}$  & $6.0\,10^8$  & $\approx$-0.8 \\
Tol~3     &  10042-2941 & 5.0   & 6.5 &  44  & $5.3\,10^{42}$  & $5.9\,10^6$$^c$  & -- \\
NGC~3256  &  10257-4338 & 94.6  & 121.6 &  45  & $1.9\,10^{45}$  & $1.3\,10^9$$^g$  & -- \\
Haro~2    &  10293+5439 & 4.8   & 5.5 &  47  & $1.6\,10^{43}$  & $6.1\,10^7$$^b$  & -0.8 \\
Mrk~36    & F11022+2924 & 0.2   & $\le$0.7 &   --  & $\le1.4\,10^{41}$  & $1.5\,10^6$$^b$   & -- \\
NGC~4214  &  12131+3636 & 14.5  & 25.5 &  38 & $5.3\,10^{42}$  & $3.6\,10^7$  & $\approx$-0.4 \\
IZw~36    &  --         &  --   & --   &   --  & $<9.4\,10^{40}$  & $2.5\,10^6$$^d$   & $>$-0.3 \\
NGC~4670  &  12428+2724 & 2.4   & 4.0  &  40 & $3.1\,10^{42}$  & $1.5\,10^7$$^e$  & -- \\
Mrk~59    &  12566+3507 & 1.8   & 2.4  &  44  & $2.0\,10^{42}$  & $4.8\,10^6$$^a$  & -0.6 \\
NGC~5253  &  13370-3123 & 31.2  & 29.8 &  51  & $4.0\,10^{42}$  & $1.0\,10^7$$^f$  & -0.1 \\
NGC~5471  &  14027+5438 & 1.8   & 2.5  &  42  & $6.0\,10^{41}$  & $4.0\,10^6$$^a$ & -0.2 \\
IIZw~70   &  14489+3546 & 0.7   & 1.5  &  36  & $2.1\,10^{42}$  & $1.3\,10^7$$^d$  & -0.3 \\
IC~4662   &  17422-6437 & 8.3   & 11.8 &  42  & $1.1\,10^{42}$  & --  & -- \\
\hline
\end{tabular} 
\end{flushleft} 
\label{rfir} 
References: $^a$ Sramek \& Weedman (1986); $^b$ Klein et al. (1984); $^c$ van Driel et al. (1991); 
$^d$ Wynn-Williams \& Becklin (1986); $^e$ Huchra et al. (1983); $^f$ Beck et al. (1996); 
$^g$ Forbes \& Ward (1993) 
\end{table*}

We took the standard 2-dimensional line-by-line spectra distributed by the
IUE observatory and reduced them using our own procedures. The steps involved
were essentially defect pixels removal (hot pixels, cosmic rays, reseau
marks), linear interpolation between adjacent pixels, background
subtraction and photometric calibration.  The accuracy of the IUE
photometric calibration at low resolution is approximately 10\%, consistent
with the optical data accuracy.  
Spectra were shifted to set the red wing of the \ion{C}{iv} absorption line
at its nominal wavelength.   Optical redshifts were not used, since the 
large size of the entrance slit can imply a
spectral shift of a few \AA\ if the object is not well centered into the
aperture.  When more than one spectrum were available we coadded them to
improve the signal-to-noise ratio.  The spectra were finally corrected from
galactic extinction using the Seaton (1979) law and {$E(B-V)$} color excesses
derived by Burstein \& Heiles (1984). We found no significant discrepancy
between the spectra extracted by our procedure and those 
 included in the catalogue of starburst galaxies by Kinney
et al. (1993).

We have measured equivalent widths of the \ion{Si}{iv} 1393+1403 and \ion{C}{iv} 1548+1551
absorption lines following the prescriptions of Sekiguchi and Anderson
(1987a).  The continuum was defined by two 40-\AA -wide clean intervals
located within 80~\AA\ on either side of the lines.  The absorption and the
emission features were identified as those parts below and above the
linearly interpolated continuum level, respectively.  Equivalent width
of the \ion{Si}{iv} and \ion{C}{iv} absorption lines were then derived by integrating over
the whole absorption feature.  Since the lines have been measured
consistently in the stars used in the Sekiguchi and Anderson (1987a)
calibration and in our sample this procedure should be adequate. The
continuum placement is certainly one of the largest source of systematic
error in the equivalent widths measurements.  We have estimated the errors
 by assigning various plausible continuum levels to the different
features. An additional source of error is the contamination of the \ion{Si}{iv}
and \ion{C}{iv} profiles by interstellar absorption lines, as well as by nebular
emission lines. When the terminal velocity in the \ion{C}{iv} line was higher than
3000~km/s we subtracted previously a gaussian fit to the \ion{Si}{ii}~$\lambda$1527
interstellar line. The contamination of the profiles by interstellar \ion{Si}{iv}
and \ion{C}{iv} absorption lines is expected to be weak, since most galaxies are
metal deficient and little affected by extinction.
 Finally, in NGC~2363 and IZw~36 the nebular contamination completely
prevented to measure the absorption lines. In the other galaxies the
contamination by nebular emission seemed to be negligible.  The measured
equivalent widths with the corresponding errors are listed in
Table~\ref{iue2}. (The merged IUE and optical spectra will be discussed
later and are displayed in Fig.~\ref{spec1}).

\subsection{ Far Infrared} 

Except for IZw~18 and IZw~36 all other objects were detected by the
Infrared Astronomical Satellite (IRAS) at 60 and/or 100 $\mu$m. The IRAS
angular resolution is around 2$\arcmin$ ~at 100~$\mu$m, so that the
measured fluxes include also the emission coming from surrounding regions.
This is not a problem in the case of blue compact dwarf galaxies, in which
star-forming regions clearly dominate over the entire galaxy.  On the other
hand, the measured flux will be strongly contaminated by emission unrelated
to the star-forming episode in more extended objects, like Mrk~710.

The  FIR parameter defined by Helou et al.  (1988) is a convenient
representation of the far infrared flux measured by IRAS from any source
dominated by thermal emission peaking between 50 and 100$\mu$m, as is
generally the case in star-forming regions.  FIR  estimates the flux
measured between 42.5 and 122.5~$\mu$m and is defined as

$$FIR = 1.26~10^{-14}[2.58f_\nu(60\mu m) +  1.00f_\nu(100\mu m)] 
{\rm ~Wm}^{-2}$$

where $f_\nu(60\mu m)$ and $f_\nu(100\mu m)$ are the nominal flux
densities.  The total far infrared flux has been estimated following the
calibration of Helou et al.  (1988): first we derived a mean dust
temperature from the $f_\nu(60\mu m)/f_\nu(100\mu m)$ ratio, assuming a
dust emissivity index of 1.  The total far infrared flux has then been
extrapolated to the 1--1000 $\mu$m band assuming a blackbody distribution.
This estimation of the total far infrared flux is a lower limit 
 since other cooler and/or hotter components have been neglected.
Puget \& L\'eger (1989) have shown for example that nearly 30\% of the
Galactic far infrared flux is emitted below 40~$\mu$m by very small grains
and large aromatic molecules, which contribute significantly to the UV
extinction.  Unfortunately, very few among our galaxies  have been
detected at 12 and 25~$\mu$m, so that a two-temperature model is not
generally possible.  Mas-Hesse (1992) found analyzing a larger sample
of BCG that in average 30\% of the far infrared flux is emitted below
40~$\mu$m, in good agreement with the results of Puget \& L\'eger  (1989).
While the far infrared luminosity we have derived is only a lower limit the
real value should not be therefore higher by much more than a factor 1.5.
We list in Table~\ref{rfir} the estimated total far infrared luminosities.

\subsection {Radio} 

Mrk~71, Mrk~710, NGC~4214 and NGC~4670 were observed in July-August 1989
with the Nan\c cay radiotelescope in the continuum at 21, 18 and 9~cm.
Only Mrk~710 and NGC~4214 were detected at 21 and 18~cm, getting upper
limits at 9~cm.  Observations were performed by drifting, i.e.,
positioning the chariot with the antenna and keeping it fixed during source
transit.  Thanks to the length of the available tracks this procedure could
be repeated every 2 minutes around 30 times each day.  The data were
processed at Nan\c cay using the SIR package.  We first integrated the
signal received during all the cycles, rejecting those affected by
external contamination (storms,...) and adding the four channels provided
by the autocorrelator (2 for vertical and 2 for horizontal polarization).
The baseline was then fitted by a polynomial and subtracted. A gaussian was
finally fitted to the remaining signal and the total flux was
evaluated. Two reference sources were observed for absolute calibration,
C287 and C295 (K\"uhr et al., 1981).  The measurements are summarized in
Table~\ref{nancay}. The accuracy of the photometric calibration lies
between 10\% at 21~cm and 20\% at 9~cm. The quoted errors have been derived
by assuming different baseline levels.

\begin{table}
\caption[]{Radio data taken at Nan\c cay radiotelescope. The fluxes are
given in mJy. Values within parentheses correspond to the reference 
fluxes of K\"uhr et al., 1981).  }
\begin{flushleft}
\begin{tabular}{lccc}
\hline
Object     & $S_{21 cm}$ & $S_{18 cm}$  &$ S_{9 cm}$  \\
\hline
C287     &  6.2 (7.1)  &  6.3  (6.3)  & 3.2 (4.1)  \\
C295     &  24.9 (22.0) & 21.0 (21.0) & 11.3 (10.0) \\
Mrk~710  &  150$\pm$10  &  126$\pm$15 & $\le$70  \\
NGC~4214 &  70$\pm$25   &  65$\pm$25  & $\le$50  \\
NGC~4670 &  $\pm$18     &  --         & $\pm$30  \\
\hline
\end{tabular} 
\end{flushleft} 
\label{nancay} 
\end{table}

Radio continuum data for the rest of the sample have been taken from the
literature. Fluxes at 6~cm are listed in Table~\ref{rfir}
together with the radio spectral index. We stress again that these radio
fluxes generally correspond to the integrated galaxy and not only to the
star-forming regions.

\section{Evolutionary synthesis models}

Our evolutionary population synthesis models have been discussed in detail
in Arnault et al. (1989), Mas-Hesse \& Kunth (1991a) and Cervi\~no \&
Mas-Hesse (1994).  They are based on the set of stellar evolutionary
tracks provided by the Geneva group (Schaller et al. 1992; Schaerer et
al. 1993ab; Charbonnel et al. 1993), which have been
computed for several metallicities between Z=2{Z$_\odot$} and Z={Z$_\odot$} /20,
including overshooting and taking into account the effect of the WR
atmospheres on the emitted spectral energy distribution. Our models
consider a stochastic IMF generated by the Monte Carlo method with initial
masses between 2 and 120 {M$_\odot$} and a power law  dependence on mass
$\phi(m)= {\rm d}N/{\rm d}m \sim m^{-\alpha}$. We have assumed three values
for the slope, the Salpeter's one, $\alpha = 2.35$, and two extreme values,
$\alpha = 1,3$.  We have used very short time steps ($\geq$ 0.05~Myr) to
account for very rapid stellar phases such as the WR one.  Two extreme star
formation regimes have been considered: an instantaneous burst (IB) and an
extended episode (EB) at a constant star formation rate.  
Calculations have been performed for a period of only 20~Myr, since we were 
mainly
interested  in the evolution of the most massive stars. After this
time, a constant star formation process enters an asymptotic phase in which
the population of massive stars (M$\geq 12~${M$_\odot$} ) is nearly constant.
 Our EB models should not be confused with those assuming constant
star formation rates over several Gyr, since even at 20 Myr stars with
masses below around 10~Myr have not yet left the Main Sequence.

The synthesis has been performed in two steps.  We have first synthesized
the population of massive stars in a young cluster as a function of age,
IMF and metallicity.  More precisely, we have calculated the number and
luminosity (per unit solar mass formed) of stars of each spectral type (O3
to M5) and luminosity class (I, III and V) by mapping the HR diagram as the
stars evolve along the evolutionary tracks.  We have then computed a set of
observational parameters covering the UV to radio range using a compilation
of spectra and fundamental properties of standard stars of each
type. Although this compilation is based only on solar metallicity stars,
we have selected a set of observational parameters which, in principle, are
not seriously affected by the metallicity of the corresponding stars. Our
dependence on metallicity is then based solely on its effects on the
stellar evolutionary tracks, which originate very different stellar
populations at a given time. We have discussed in Mas-Hesse \& Kunth
(1991a) the limitations of this approach.  Synthesis models based on
stellar atmosphere models at different metallicities have been presented in
the last years (Garc\'{\i}a-Vargas et al.  1994; Schaerer \& Vacca
1998). While these models provide a more accurate prediction of the shape
of the ionizing continuum (and are thus better suited to synthesize the
emission line spectrum), the parameters we have synthesized are not
significantly affected.  In the following paragraphs we will discuss
briefly our set of parameters and their validity for constraining the
properties of the star formation processes:

\subsection{Ionizing flux and effective temperature}

The number of ionizing photons can be measured from the luminosity of the
Balmer emission lines and also from the thermal radio emission. There are
essentially two technical problems: first, the extinction affecting the
nebular gas in global has to be well determined and corrected. As we will
see later, inhomogeneities in the dust particles distribution can make this
task rather complex. It is also very difficult to account for the number of
high energy photons directly absorbed by dust, which do not contribute to
the ionization process. We have assumed as a first approximation that in
average 30\% of the photons below 912~\AA\ are directly absorbed, as found
in Galactic \ion{H}{ii} regions by Smith et al. (1978) and Mezger (1978), but the
exact value will depend on each galaxy. Second, it is needed to integrate
the emission line fluxes over the whole emitting region. If such region is
ionization bounded and the flux is spatially integrated over the whole
emitting region, the complete number of ionizing photons can be
estimated. If this is not the case, a significant fraction of the photons
could escape without being accounted for when comparing with 
  models predictions. As explained above, we have measured {$L(H\beta)$}\
through a large slit in order to cover essentially the whole ionized regions
and minimize these problems.

Effective temperatures provide an additional constraint on the mass
distribution of stars that dominates the ionizing emission of the cluster. The
effective temperature can not be directly measured but has to be estimated
from its effects on several emission lines. Unfortunately, these effects
are also very dependent on other properties of the nebular gas (filling
factor, density, metallicity,...), which not always can be well determined.
Effective temperature estimates therefore strongly depend on the
photoionization scenario (see V\'{\i}lchez \& Pagel 1988 and Cervi\~no \&
Mas-Hesse 1994). The determination of the effective temperature can be
in error if  spatial integration of
the emission flux over the whole region is not performed.

\subsection{Spectral energy distribution (SED)}

We have computed the spectral energy distribution in the UV, optical, near
and far infrared and radio ranges. Each of these energy ranges can be
dominated by different stellar populations, so that the comparison of the
whole predicted and observed SEDs is a powerful method to set constraints
on the structure of the stellar population. While the UV emission is
dominated by young, hot stars, the principal contributors to the near
infrared bands are red supergiant stars formed during the burst (or even
nebular continuous emission, depending on the evolutionary state of the
burst).  The radio continuum is essentially due to thermal emission when
the burst is young, but can be dominated by the non-thermal emission of
supernova remnants once the most massive stars have ended their
lifetime. The principal problem is the contamination of the observed flux
by an underlying population of older stars, unrelated to the present
burst. While this problem is minimal in the UV range, it can seriously
affect the optical and near infrared ranges, especially if an important
population of red giants is present (Thuan 1983). We have minimized this
effect  by selecting objects with very strong UV
continuum, expected to be dominated in most wavelengths by the emission
from young massive stars. We will see later that, nevertheless, in some
cases the contribution of stars not related to the burst is still 
significant. 

\begin{figure}
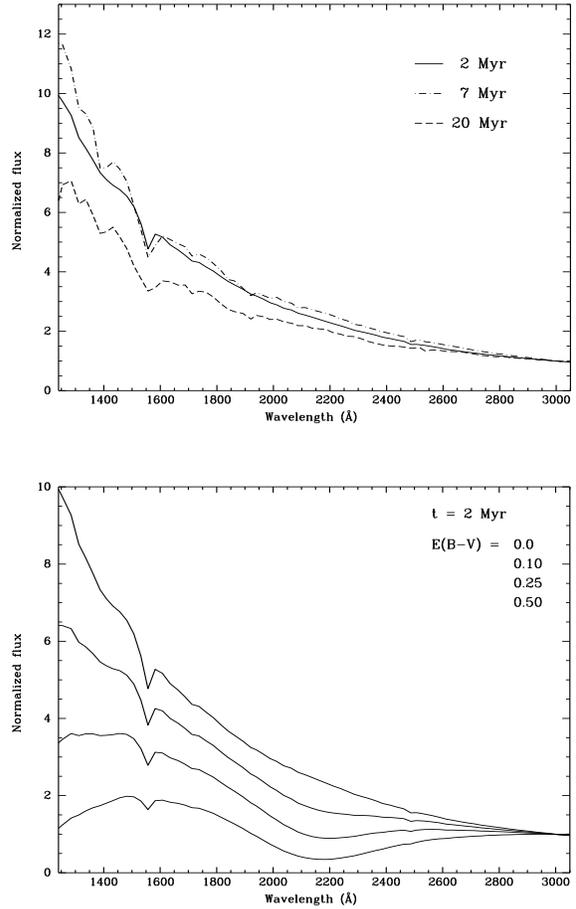
 
\begin{center}\mbox{\epsfxsize=8cm \epsfbox{ms8377.f2a}}\end{center}
\begin{center}\mbox{\epsfxsize=8cm \epsfbox{ms8377.f2b}}\end{center}
\caption[]{
a) Synthesized continuum spectra for different ages
between 2 and 20~Myr. It can be seen that the UV slope varies  weakly
in the first 7~Myr, becoming even bluer than at zero age. 
b) Effects of interstellar extinction on a synthetic
continuum of a 2~Myr old cluster. 
Even low amounts of dust modify significantly the shape of the
UV continuum. 
}
\label{cext}
\end{figure}

Constraints that can be derived from the optical-UV SED are hampered by
two facts: first, the UV slope varies very weak\-ly with age during the first
10 Myr (Fig.~\ref{cext}a). Second, even low amounts of interstellar dust
produce a significant reddening of the UV emission (Fig.~\ref{cext}b).  The
energy these dust particles absorb is reradiated thermally in the far
infrared range, as computed in Mas-Hesse \& Kunth (1991a). Therefore, by
simultaneously analyzing the shape of the UV continuum and the far infrared
emission we can estimate the amount of extinction and recover the original
SED. As a by-product we derive important information about the shape and
strength of the extinction law in these objects, as we will discuss
later. It is thus crucial to build information from a wide wavelength range to
properly derive the properties of the stellar population  in the
young clusters.

An additional problem when comparing the FIR and radio luminosities is
 that both have been generally obtained
through very large apertures, covering essentially the whole galaxies. The
comparison is only  meaningful  in compact galaxies, in which the
contribution in these bands is dominated by the star--forming region.

\begin{figure}
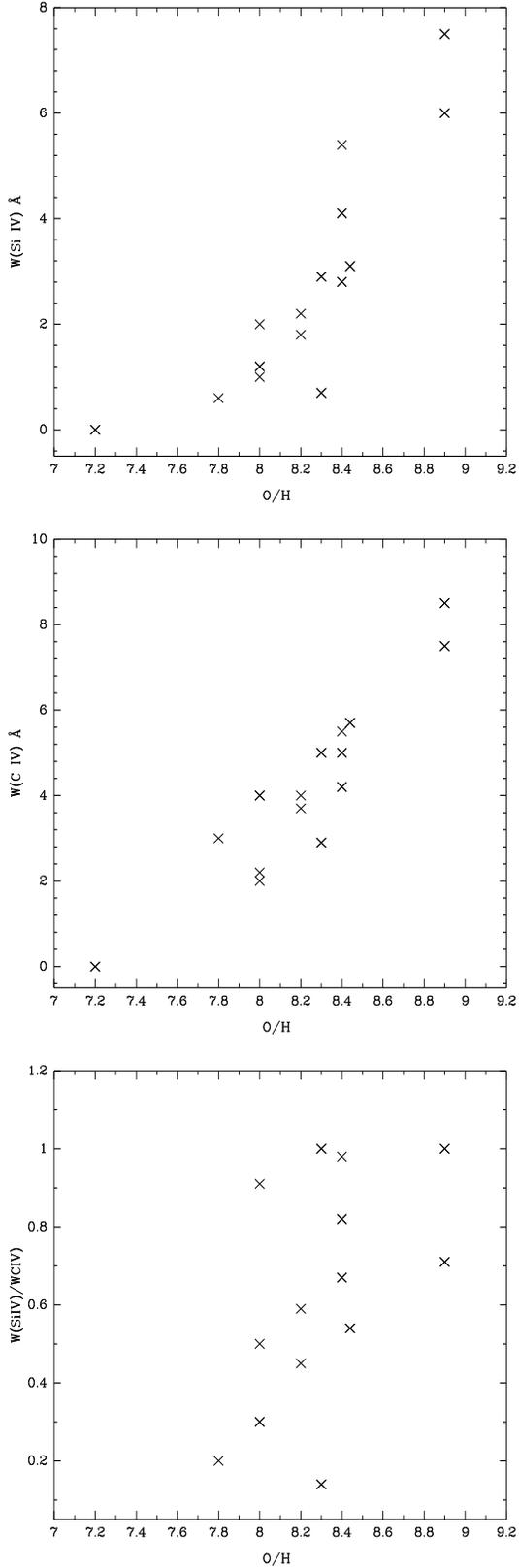
 
\begin{center}\mbox{\epsfxsize=7cm \epsfbox{ms8377.f3a}}\end{center}
\begin{center}\mbox{\epsfxsize=7cm \epsfbox{ms8377.f3b}}\end{center}
\begin{center}\mbox{\epsfxsize=7cm \epsfbox{ms8377.f3c}}\end{center}
\caption[]{
a) {$W(\ion{Si}{iv})$}, b) {$W(\ion{C}{iv})$} and c) {$W(\ion{Si}{iv})/W(\ion{C}{iv})$}  vs. the O/H abundance
as measured from nebular emission lines for the objects in the sample.
}
\label{wsc}
\end{figure}

\subsection{Ratio of the \ion{Si}{iv} $\lambda$1400 and \ion{C}{iv} $\lambda$1550
absorption lines equivalent widths}

The \ion{Si}{iv} $\lambda$1400 and \ion{C}{iv} $\lambda$1550 absorption lines provide a
direct evidence for the presence of a massive stellar population.  They
frequently show a P~Cygni profile, as seen in the UV spectra of individual
stars in our Galaxy. Their ratio is very sensitive to the spectral type of
stars dominating the cluster emission since it varies by more than a factor
10 between O6 and B1 and shows a significant luminosity effect (Sekiguchi
\& Anderson 1987a; Mas-Hesse \& Kunth 1991a).  {$W(\ion{Si}{iv})/W(\ion{C}{iv})$} can then provide a
first estimate about the dominant stellar population independently on any
evolutionary model assumption. The use of the equivalent width of these
lines to derive information about the stellar content of the clusters was
first introduced by Sekiguchi \& Anderson (1987a+b). Mas-Hesse \& Kunth
(1991a) and Cervi\~no \& Mas-Hesse (1994) have synthesized the
 {$W(\ion{Si}{iv})/W(\ion{C}{iv})$} ratio
as a function of age, IMF and metallicity of the star--forming
region. Robert, Leitherer and Heckman (1993) have synthesized the whole
profiles of the lines for solar metallicity bursts based on a library of
high resolution IUE spectra.

A major problem is that these equivalent widths strongly depend on the
metallicity of the gas from which the stars formed. Profile fitting
techniques or even approaches based on the analysis of single lines require
the availability of a library matching the metallicity of the object under
study. On the other hand, the ratio of both equivalent widths,
 {$W(\ion{Si}{iv})/W(\ion{C}{iv})$},
should be essentially independent on metallicity.  We have plotted in
Figs.~\ref{wsc}a,b,c the measured {$W(\ion{Si}{iv})$}, {$W(\ion{C}{iv})$} and 
{$W(\ion{Si}{iv})\-/W(\ion{C}{iv})$} versus the nebular
oxygen abundance for the objects in our sample.  It can be seen that the
upper envelope of the {$W(\ion{Si}{iv})$} and {$W(\ion{C}{iv})$} vs. O/H distributions shows a clear
trend toward lower strengths of \ion{Si}{iv} and \ion{C}{iv} absorption lines at low
metallicities. The dispersion below this upper envelope is affected by
evolutionary effects, since the lines become weaker when the cluster
evolves. But the trend is weaker or inexistent at all in the case of the
{$W(\ion{Si}{iv})/W(\ion{C}{iv})$} ratio, as shown in Fig.~\ref{wsc}c.  Leitherer \& Lamers (1991) have
argued that the {$W(\ion{Si}{iv})\-/W(\ion{C}{iv})$} ratio could nevertheless be a function of the
intrinsic metallicity of the stars. The effect should be negligible down to
Z = {Z$_\odot$} /3, but they predict an increase of the ratio at lower
metallicities.  Indeed, the \ion{Si}{iv} line is expected to be a saturated
photospheric line and therefore insensitive to Z while the \ion{C}{iv} line would
originate in the wind and therefore decrease with decreasing
Z. Figs.~\ref{wsc}a,b show that both lines are similarly sensitive to
metallicity, contrary to predictions. In fact a clear upper envelope is
well defined for both absorption lines, especially for \ion{C}{iv}, which is
generally stronger. This envelope defines the relative abundance of carbon
and silicon with respect to oxygen in the gas from which the stars
formed. A similar correlation between {$W(\ion{Si}{iv})$} and {$W(\ion{C}{iv})$} and the nebular oxygen
abundance has been found by Storchi-Bergmann, Calzetti \& Kinney (1994) in
a sample of 44 star-forming galaxies.

As we pointed out in section 2, both the \ion{Si}{iv} and \ion{C}{iv} absorption lines of
photospheric origin can be further contaminated by the additional
contribution of interstellar absorptions adding a source of uncertainty in
the determination of the {$W(\ion{Si}{iv})/W(\ion{C}{iv})$} ratio. In any case, the calibration of the
{$W(\ion{Si}{iv})/W(\ion{C}{iv})$} ratio with spectral types already includes some degree of
contamination by the Galactic ISM. HST allows to test to which extent the
contribution by interstellar lines can dominate the {$W(\ion{Si}{iv})/W(\ion{C}{iv})$} ratio.  From the
comparison of the NGC~4214 HST spectra with synthetic ones presented by
Leitherer et al. (1996), we can see that the effect of \ion{Si}{iv} and \ion{C}{iv}
interstellar absorptions on the integrated equivalent widths of the P~Cygni
profiles is small and of the same order (or weaker) than the uncertainties
induced by noise and by the placement of the continuum. We have therefore
neglected this effect and assumed it is contained within our error bars.

\subsection{{H$\beta$}~equivalent width}

{$W(H\beta)$} has been recognized for a long time as a very useful parameter to
constrain the elapsed time  since the onset of a star formation episode
(Copetti et al. 1985; Melnick et al. 1985). Since it relates the number of
the most massive stars ($ M > 20$~{M$_\odot$} ) to lower mass ones it decreases
continuously with time as the most massive stars cease to generate ionizing
photons. Nevertheless, this age effect can be mixed with those induced by
variations in the slope and upper mass limit of the IMF.  Therefore, it is
needed to compare simultaneously different parameters to constrain both the
age and the IMF.  This can be achieved for example by using the
 {$W(\ion{Si}{iv})/W(\ion{C}{iv})$} vs.
{$W(H\beta)$} diagrams presented in Mas-Hesse \& Kunth (1991a) and updated by
Cervi\~no \& Mas-Hesse (1994), as we will discuss later.

While {$W(H\beta)$} is in principle easy to measure on optical spectra there are a
number of technical problems than can lead to erroneous results. First of
all, {$W(H\beta)$} has to be measured over the complete region comprising all the
stars that contribute to the continuum and also the whole {H$\beta$} emitted
flux. Since the extension of the ionized nebular gas is generally much
larger than the region where the continuum is emitted, {$W(H\beta)$} can easily be
underestimated. On the other hand, within a large aperture the
contamination of the continuum by older stars might become
important. Finally, the extinction may affect differently the nebular
emission and the stellar continuum hence the determination of
{$W(H\beta)$}. We will discuss later that in some cases the continuum seems to be
significantly less reddened than the emission lines, contrary to the usual
assumption that no such correction has to be applied to {$W(H\beta)$}.

\begin{figure}
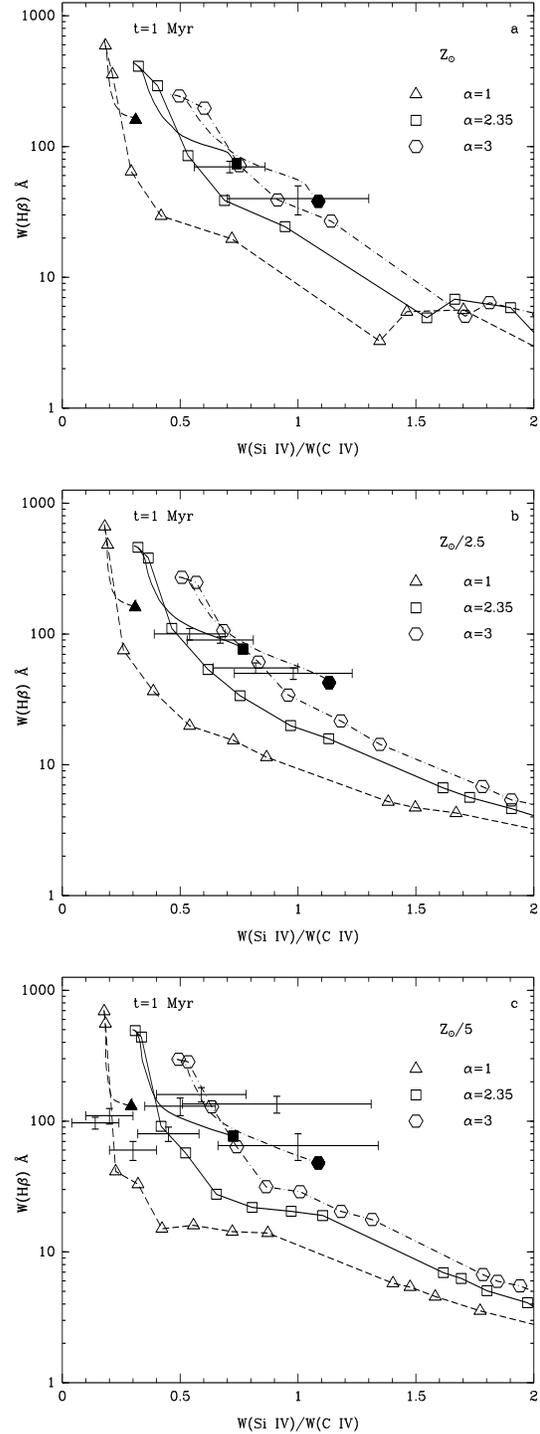
 
\begin{center}\mbox{\epsfxsize=7cm \epsfbox{ms8377.f4a}}\end{center}
\begin{center}\mbox{\epsfxsize=7cm \epsfbox{ms8377.f4b}}\end{center}
\begin{center}\mbox{\epsfxsize=7cm \epsfbox{ms8377.f4c}}\end{center}
\caption[]{
Diagnostic diagrams {$W(H\beta)$} vs. {$W(\ion{Si}{iv})/W(\ion{C}{iv})$} ratio for different
metallicities. The tracks have been computed for three IMF slopes, with a
tick every 1~Myr. Solid symbols correspond to the asymptotic predictions
for continuous star formation processes at 20~Myr. 
The observational values from the objects in our sample have been included
with estimated error bars. 
}
\label{whws}
\end{figure}

\subsection{The Wolf-Rayet bump strength}

Wolf-Rayet stars are considered to be highly evolved descendants of the
most massive O-stars. They are extreme Population~I stars and have spectra
characterized by broad emission lines resulting from dense, high-velocity
winds. These stars are detectable in external galaxies by their prominent
emission lines at around 4650-4690~\AA\ (the ``Wolf-Rayet bump''). This
bump has been detected in many emission-line galaxies (Allen et al. 1976;
Kunth \& Sargent 1981, Kunth \& Schild 1986; Conti 1991, Conti \& Vacca
1992; Schaerer et al. 1999), providing a strong evidence that WR stars are
indeed the descendants of massive O-type stars. Arnault et al. (1989),
Cervi\~no \& Mas-Hesse (1994), Meynet (1995) and Schaerer \& Vacca (1998)
have discussed the dependence of the WR bump strength on the parameters
that define the star--forming episode (metallicity, age, IMF
slope,...). The two most interesting properties of this feature are its
strong dependence on metallicity and the constraints it can impose on the
age of the cluster. Since the WR phenomenon is tightly coupled to the
generation of strong stellar winds, its incidence decreases significantly
with metallicity, so that at Z~=~{Z$_\odot$} /20 only very massive stars (initial
mass $>$ 80 {M$_\odot$} ) might become WR. This small mass range implies that the
detection of the WR bump in low metallicity galaxies can provide excellent
on the upper mass limit of the IMF.

The relative population of WR over O stars is usually measured through the
{$L(WR)/L(H\beta)$} ratio. To compare with mo\-del predictions it is again necessary to 
integrate over the whole ionized region.  The
observational measurements of this ratio might also be strongly affected by
differential extinction. Since $L(WR_{bump})$ is of stellar origin, it
should be affected by the same extinction than the stellar continuum. On the
other hand, $L(H\beta)$ is of nebular origin and might suffer from
a larger amo\-unt of extinction. Ignoring this effect may lead to a
significant overestimation of the {$L(WR)/L(H\beta)$} ratio.

Finally, recent results by Cervi\~no (1998) show that if a
significant fraction of massive stars are formed in binary systems, mass
transfers episodes can lead to the formation of WR stars during longer
periods of time than predicted by models based on the evolution of single
stars alone. The age calibration through the WR features has to be therefore
taken with caution. 

\section {Model fitting}

It is clear from the above discussion that no single observable taken alone can
constrain the properties of  star forming processes. Only by
simultaneously fitting several parameters over a large wavelength range 
tight to different physical proce\-sses can we optimize the unicity of the
results. To fit the predictions of the evolutionary synthesis models to the
observational data we rely upon the following steps (additional details on 
the fitting technique are given in Mas-Hesse \& Kunth 1991a):

\begin{itemize} 

\item[i.-] The diagnostic diagrams {$W(H\beta)$} vs.
 {$W(\ion{Si}{iv})/W(\ion{C}{iv})$} (Fig.~\ref{whws}) 
provide a first constraint on both the IMF slope and the age of the
cluster (Cervi\~no \& Mas-Hesse, 1994).
 Since they are metallicity dependent, we have used the O/H nebular
abundance as an input value. Although the evolution of instantaneous or
extended star formation episodes is significantly different, it is
generally not possible to discriminate between both modes from these
diagrams alone. For galaxies with no available {$W(\ion{Si}{iv})/W(\ion{C}{iv})$} ratio
 we assumed a Salpeter's IMF and used {$W(H\beta)$} for constraining
the age.

\item[ii.-] The age has been further constrained by 
 the Wolf-Rayet bump. The {$L(WR)/L(H\beta)$} ratio is sensitive to the star
formation regime (instantaneous or extended).  On the
other hand, when available, the radio spectral index has been used to
disentangle between instantaneous or extended bursts. As discussed in
Mas-Hesse \& Kunth (1991a), the number of ionizing photons emitted in a
burst in which new massive stars are being continuously formed is so high,
that radio emission is always dominated by the thermal contribution from
nebular gas, while the non-thermal fraction associated to supernova
remnants is almost negligible. The detection of radio spectral indexes
below $\alpha = -0.5$ allows to reject extended star formation
episodes. Since the radio emission is integrated over the whole galaxy,
this constraint is reliable only for small objects dominated by the
starburst, but not  for larger galaxies with  important disk
population.

\item[iii.-] We have fitted the UV continuum by  models
 normalized at 3000 \AA . We have selected 11 bands between 1285
and 3000 \AA\ chosen to be relatively free from emission or absorption
lines to perform the fitting. When only the 1200-1900 \AA\ IUE range was
available, 8 points were selected and the normalization was done at 1930
\AA . The fitting has been performed assuming three different extinction
laws: Galactic (Cardelli et al. 1991), Large Magellanic Cloud (LMC) (Nandy
et al. 1981) and Small Magellanic Cloud (SMC) (Pr\'evot et al. 1984).
The fit is qualified by the $\epsilon$ parameter, which measures the deviation
between the synthetic and the observed averaged  continua.

 After these first three steps of the fitting procedure
we ended with a relatively narrow range of valid models, as well as with an
optimal extinction law and the corresponding value of the {$E(B-V)$} color
excess.

\item[iv.-] Comparing the observed luminosity at 3000 (1930) \AA\ with the one
predicted by the best models we derived the total amount of gas transformed
into gas since the beginning of the burst (predictions
are normalized to 1 {M$_\odot$} ).  This mass corresponds to stars within the IMF
boundaries  (2-120 {M$_\odot$} and the corresponding slope). Once this
normalization factor has been derived, we have computed model predictions
for absolute parameters, like {$L_{FIR}$} and $L(H\beta)$.  {$L_{FIR}$}, which is a
function of the total reddening affecting the intrinsic UV SED, 
has then been compared to the observational value as a consistency
check of the assumed extinction correction. A final constraint is provided
by comparing observed and predicted {H$\beta$} luminosities.  When available we
have used {H$\beta$} luminosities observed through apertures similar to the IUE
ones. 

\item[v.-] We have finally compared  predicted and observed optical
continua (only for those objects observed through large apertures). While
in a burst dominated object the contribution to the optical continuum by an
older underlying population should be negligible, this contribution can be
important in some galaxies. Subtracting the predicted from the observed
continuum flux we have estimated this contribution and have corrected the
measured {$W(H\beta)$} value from the contamination by underlying population. With
this corrected value of {$W(H\beta)$} we have iterated again at step i.

\end{itemize} 

\begin{figure*} 
\begin{center}\mbox{\epsfxsize=16cm \epsfbox{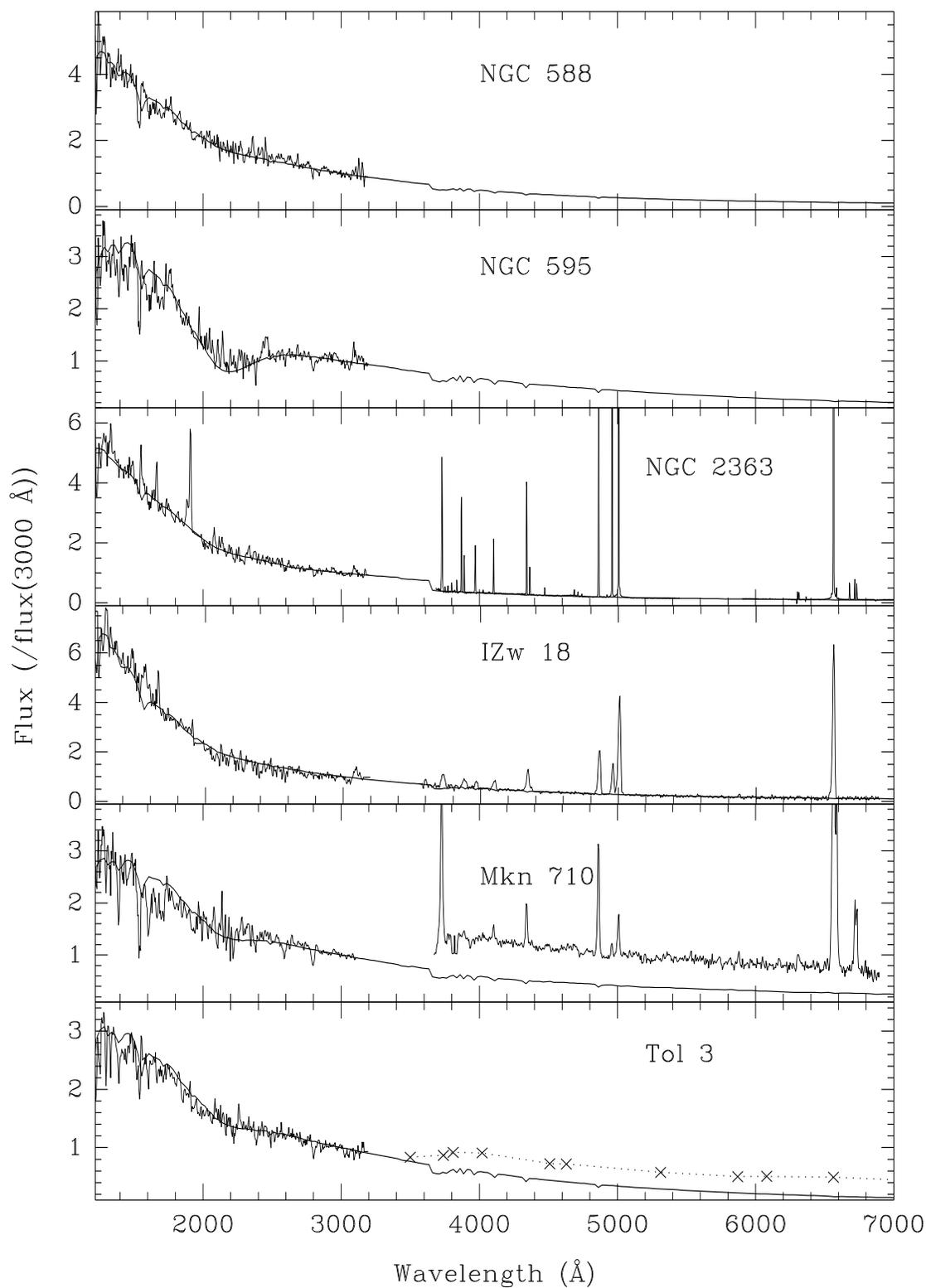}}\end{center}
\caption[]{
Synthetic over observational spectra of the different
galaxies in the sample. Crosses and dotted lines correspond to the optical
spectroscopy taken from Storchi-Bergmann et al. (1995). In
those cases with optical spectra taken through narrow slits (see the text),
the optical continuum has been normalized to the predicted flux. 
The optical range from NGC~5471 is a FOS spectrum provided by
Rosa \& Benvenuti (1994). The optical spectrum of NGC~2363 has been
provided by R. Gonz\'alez-Delgado and E. P\'erez.   
}
\label{spec1}
\end{figure*}

\addtocounter{figure}{-1}

\begin{figure*} 
\begin{center}\mbox{\epsfxsize=17cm \epsfbox{ms8377.f52}}\end{center}
\caption[]{{\em  Continuation.}}
\end{figure*}

\addtocounter{figure}{-1}

\begin{figure*} 
\begin{center}\mbox{\epsfxsize=17cm \epsfbox{ms8377.f53}}\end{center}
\caption[]{{\em  Continuation.}}
\end{figure*}

\section {Results and discussion}

We have summarized in Table~\ref{res} the results of our evolutionary
synthesis models applied to 17 star--forming regions.  We list in this
table the parameters that provide the best fit, together with their
corresponding predicted and measured values. The mass of gas transformed
into stars corresponds in all cases to the whole IMF range between 2 and
120 {M$_\odot$}. This value does not necessarily correspond to the total amount
of gas transformed into stars, since a significant fraction of gas is
expected to have formed lower mass stars, not included in our
computations. On the other hand, if the IMF is truncated at upper mass
limits lower than 120~{M$_\odot$}, the total mass involved would be
correspondingly lower.  We can see from the table that in general a
solution can be found which simultaneously reproduces the whole set of
multiwavelength parameters. Nevertheless, as we will discuss later, in some
individual cases either the solution is not unique or some of the
parameters can not be properly reproduced by the models.

\begin{figure} 
\begin{center}\mbox{\epsfxsize=7cm \epsfbox{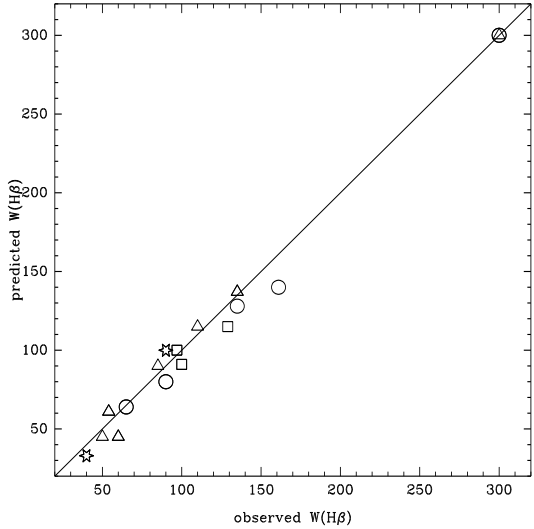}}\end{center}
\caption[]{
Predicted vs. observed {$W(H\beta)$} values in \AA. Squares: G\ion{H}{ii} regions; circles:
irregulars; triangles: blue compacts;
stars: spiral and merger.
}
\label{rwhb}
\end{figure}

The observational values of {$W(H\beta)$} and {$L(WR)/L(H\beta)$} have been corrected from the
contribution to the optical continuum of underlying older stars, as well as
from differential extinction between continuum and emission lines, es
explained below. We have plotted in Figs.~\ref{spec1} the continuum
corresponding to the best fitting models over the observed IUE and optical
spectra of the different objects.  In Figs.~\ref{rwhb},\ref{rwswc} and
\ref{rhb} we show the comparison between the predicted and observational
values of {$W(H\beta)$}, {$W(\ion{Si}{iv})/W(\ion{C}{iv})$} and {$L(H\beta)$}.

\begin{table*}
\caption[]{Predicted vs. observational parameters. We have included here
 model predictions that best fit a maximum number of observable
parameters. Observational data are given in the second line of the 
corresponding entries. Note that the observational {$W(H\beta)$} and {$L(WR)/L(H\beta)$} values
have been corrected as explained in the text. The
$L(4860)_{sb}$/$L(4860)_{tot}$ ratio measures the fraction of the continuum 
at 4860~\AA\ originated by the young stars over the total continuum within
the aperture, as explained in the text. When both Extended and Instantaneous 
Burst models could be fitted, the best fitting one has been considered, and
the age derived from the second one has been included within parentheses.  }
\begin{flushleft}
\begin{tabular}{@{}l@{}ccccccccc@{}}
\hline
&        NGC~588 & NGC~595 & NGC~2363 & IZw~18 & Mrk~710&Tol~3&NGC~3256&Haro~2&Mrk~36\\
\hline
Metallicity ($Z$)  & 0.004 &0.008 & 0.001 &0.001&0.020&0.004&0.020&0.008&0.001\\
IMF slope & -1 &-2.35&-2.35&-2.35&-2.35&-2.35&-3&-1\\
Age (Myr) & 2.8 &3.1&2.6&13(3.0)&10(4.0)&3.0&4.5&4.8&2.9\\
SFR & IB &IB&IB&EB(IB)&EB(IB)&IB&IB&IB&IB\\
\hline
{$E(B-V)$}$_{UV}$ & 0.11 &0.29&0.08&0.05&0.19&0.16&0.16&0.12&0.05\\
Extinction law   & LMC &Gal&LMC&Gal&LMC&LMC&SMC&LMC&SMC\\
{$E(B-V)$}$_{opt}$ &  0.10 & 0.33 & 0.16 & 0.05 & 0.29 & 0.17 & 0.85 & 0.22 & 0.0 \\
$L_{UV}^{synt}$  &
 1.81\,10$^{33}$ &1.91\,10$^{32}$&4.14\,10$^{32}$&2.78\,10$^{32}$&1.39\,10$^{32}$&3.75\,10$^{32}$&
   1.22\,10$^{32}$ &9.90\,10$^{31}$&7.38\,10$^{33}$\\
(erg s$^{-1}$ \AA$^{-1}${M$_\odot$}$^{-1}$)&&&&&&&&&\\
$M_{trans}$ ({M$_\odot$})  & 
 5.34\,10$^{2}$ &5.01\,10$^{3}$ &4.69\,10$^{4}$&1.33\,10$^{5}$&1.42\,10$^{6}$&5.92\,10$^{5}$&
   9.84\,10$^{7}$ &6.62\,10$^{6}$&9.23\,10$^{3}$\\
$L(4860)_{sb}$/$L(4860)_{tot}$ & -- & -- & 1.0 & 0.5 & 0.4 & 0.5 & 0.6 & 0.6 & -- \\  
\hline
{$W(\ion{Si}{iv})/W(\ion{C}{iv})$}  &0.2&0.5&0.3&--&0.6&0.4&1.0&1.0&0.2\\
     &0.14&0.54&--&--&0.71&0.45&1.0&0.98&0.20\\
\\
{$W(H\beta)$} (\AA)  &100&91&300&137&100&90&33&45&115\\
              &97&100&300&135&90&85&40&50&110\\
\\
{$T_{eff}$} (K) &39000&37300&42000&42000&40300&38900&33800&34500&41000\\
             &36600&35000&42700&35000&39800&38200&37000&35400&40000\\
\\
{$L(WR)/L(H\beta)$}  &0.0&0.05&0.0&0.0005&0.04&0.05&0.20&0.04&0.0\\
        &-- &--  &0.0&0.05&0.06&0.09&--&0.06&--\\
\\
{$L(H\beta)$}  (erg s$^{-1}$) &
 5.1\,10$^{37}$ & 1.1\,10$^{38}$&1.9\,10$^{39}$&1.9\,10$^{39}$&1.6\,10$^{40}$&1.5\,10$^{40}$&
   2.2\,10$^{41}$&1.8\,10$^{40}$&9.4\,10$^{38}$\\
   &1.2\,10$^{38}$&4.5\,10$^{38}$&1.6\,10$^{39}$&1.5\,10$^{39}$&1.8\,10$^{40}$&1.6\,10$^{40}$&
   1.9\,10$^{42}$ &4.2\,10$^{40}$&7.3\,10$^{38}$\\
\\
{$L_{FIR}$} (erg s$^{-1}$) &
 1.4\,10$^{40}$ &4.5\,10$^{40}$&2.8\,10$^{41}$&2.6\,10$^{41}$&5.5\,10$^{42}$&4.7\,10$^{42}$&
   1.5\,10$^{44}$ &8.9\,10$^{42}$&2.0\,10$^{41}$\\
   &$\le$1.2\,10$^{40}$&$\le$8.3\,10$^{40}$&2.7\,10$^{41}$&$<$2.6\,10$^{41}$&7.4\,10$^{42}$&5.3\,10$^{42}$&
   1.9\,10$^{45}$ &1.6\,10$^{43}$&$\le$1.4\,10$^{41}$\\
\\
{$L_{radio}$} (Jy kpc$^2$) &
1.5\,10$^{4}$   &3.5\,10$^{4}$&5.6\,10$^{5}$&4.9\,10$^{5}$&5.4\,10$^{6}$&4.5\,10$^{6}$&
   1.1\,10$^{8}$&6.3\,10$^{6}$&2.8\,10$^{5}$\\
    &--&6.3\,10$^{4}$&1.3\,10$^{6}$&2.3\,10$^{6}$&6.0\,10$^{8}$&5.9\,10$^{6}$&1.3\,10$^{9}$&
    6.1\,10$^{7}$&1.5\,10$^{6}$\\
\\
{$\alpha_{radio}$} &-0.1&-0.2&-0.1&-0.2&-0.2&-0.1&-0.2&-0.3&-0.1\\
          &-- &$\approx$0&$\approx$0&-0.12&-0.8&--&--&-0.86&--\\
\hline
\end{tabular} 
\end{flushleft} 
\label{res} 
\end{table*}

\addtocounter{table}{-1}

\begin{table*}
\caption[]{{\it Continuation.}}
\begin{flushleft}
\begin{tabular}{lcccccccc}
\hline
&        NGC~4214 & IZw~36 & NGC~4670 & Mrk~59 & NGC~5253 & NGC~5471 & IIZw~70 & IC~4662 \\
\hline
Metallicity ($Z$)  &0.008&0.001&0.008&0.004&0.004&0.004&0.004&0.004\\
IMF slope & -2.35&-2.35&-3&-3&-3&-2.35&-1&-3\\
Age (Myr) & 3.5&2.7&4.0&3.0&3.0&2.9&3.8&13(4.0)\\
SFR & IB&IB&IB&IB&IB&IB&IB&EB(IB)\\
\hline
{$E(B-V)$}$_{UV}$ &0.07&0.007&0.13&0.05&0.10&0.07&0.09&0.047\\
Extinction law   &SMC&SMC&LMC&Gal&SMC&Gal&LMC&SMC\\
{$E(B-V)$}$_{opt}$ & 0.28  & 0.0  & 0.32 & 0.0 & 0.19 & 0.15 & 0.17 & 0.09 \\
$L_{UV}^{synt}$  &5.06\,10$^{32}$&2.14\,10$^{33}$&1.11\,10$^{32}$&1.90\,10$^{32}$&1.40\,10$^{32}$&
                 5.83\,10$^{32}$&1.26\,10$^{33}$&1.19\,10$^{32}$\\
(erg s$^{-1}$ \AA$^{-1}${M$_\odot$}$^{-1}$)&&&&&&&&\\
$M_{trans}$ ({M$_\odot$})  & 2.63\,10$^{5}$&1.28\,10$^{4}$&3.03\,10$^{6}$&1.49\,10$^{6}$&8.00\,10$^{5}$&
                 1.17\,10$^{5}$&2.17\,10$^{5}$&5.35\,10$^{5}$\\
$L(4860)_{sb}$/$L(4860)_{tot}$ & 0.5 & 1.0 & 0.5 & 1.0 & 0.5 & -- & 0.8 & -- \\
\hline
{$W(\ion{Si}{iv})/W(\ion{C}{iv})$} & 0.5&0.2&0.8&0.6&0.6&0.4&0.3&0.0\\
     & 0.67&--&0.82&0.91&0.59&0.50&0.30&1.0\\
\\
{$W(H\beta)$} (\AA) & 80&300&61&128&140&115&45&64\\
            & 90&300&54&135&160&129&60&65\\
\\
{$T_{eff}$} (K) &36700&41000&36000&38500&39000&39000&36800&40000\\
           &37000&40000&37000&40000&37000&39000&36000&38000\\
\\
{$L(WR)/L(H\beta)$}  &0.05&0.0&0.06&0.003&0.003&0.005&0.11&0.003\\
      &0.05&--&$\le$0.1&0.08&0.009&0.02&--&0.005\\
\\
{$L(H\beta)$}  (erg s$^{-1}$) &4.1\,10$^{39}$&5.1\,10$^{38}$&1.2\,10$^{40}$&1.2\,10$^{40}$&7.0\,10$^{39}$&
                        3.2\,10$^{39}$&6.8\,10$^{39}$&1.9\,10$^{39}$\\
                      &1.3\,10$^{40}$&8.6\,10$^{38}$&2.1\,10$^{40}$&1.4\,10$^{40}$&6.9\,10$^{39}$&
                        2.0\,10$^{39}$&1.4\,10$^{40}$&2.2\,10$^{39}$\\
\\
{$L_{FIR}$} (erg s$^{-1}$) &1.2\,10$^{42}$&3.7\,10$^{40}$&5.3\,10$^{42}$&2.0\,10$^{42}$&1.6\,10$^{42}$&
                        6.0\,10$^{41}$&2.6\,10$^{42}$&4.3\,10$^{41}$\\
                  &5.3\,10$^{42}$&$<$9.4\,10$^{40}$&3.1\,10$^{42}$&2.0\,10$^{42}$&4.0\,10$^{42}$&
                        6.0\,10$^{41}$&2.1\,10$^{42}$&1.1\,10$^{42}$\\
\\
{$L_{radio}$} (Jy kpc$^2$) &1.4\,10$^{6}$&4.7\,10$^{5}$&4.2\,10$^{6}$&3.5\,10$^{6}$&2.0\,10$^{6}$&
                        9.6\,10$^{5}$&2.8\,10$^{6}$&6.5\,10$^{5}$\\
                      &3.6\,10$^{7}$&2.5\,10$^{6}$&1.5\,10$^{7}$&4.8\,10$^{6}$&1.0\,10$^{7}$&
                        4.0\,10$^{6}$&1.3\,10$^{6}$&--\\
\\
{$\alpha_{radio}$} &-0.2&-0.1&-0.3&-0.1&-0.1&-0.1&-0.4&-0.2\\
        &-0.4&$\ge$-0.3&--&-0.6&-0.1&-0.2&-0.3&--\\
\hline
\end{tabular} 
\end{flushleft} 
\label{res2} 
\end{table*}

\subsection{Age and star formation regime} 

The star formation regime seems to vary from
object to object. While there are strong constraints in few objects toward
 constant star formation rates during at least 10-15~Myr,
in most cases the formation of massive stars seems to have been essentially
coeval, as found  in galaxies hosting large numbers of WR stars.
Unfortunately, evolutionary synthesis models predict similar values for
several parameters 3-5~Myr after an instantaneous burst and  after 20 Myr
of continuous star formation (Mas-Hesse \& Kunth 1991a).
 For the purposes of this paper we have assumed in
these ambiguous cases that the star formation episodes have been nearly
instantaneous.

It has been postulated in the past that  strong ionizing flux produced
by newly formed massive stars could inhibit further formation of stars,
explaining why massive stars seem to be coeval in many bursts. On the other
hand, we have found that massive stars have been continuously formed during
several million years in some galaxies. HST imaging of IZw~18, for example,
has allowed to identify individual clusters with different ages (Hunter \&
Thronson 1995). The superposition of several of these clusters at different
 ages
would mimic an extended star formation episode when analyzed globally.
  Therefore, the detection of starbursts  forming stars
during extended periods of time can still be reconciled with the idea that
individual clusters of massive stars are coeval.

The spread in the evolutionary status of the studied starbursts is rather
narrow. Their mean age is around 4~Myrs, in any case within the range
2.5-6.5~Myrs. Our sample is obviously biased toward very young bursts,
since galaxies have been selected from their strong emission lines. After
some 7~Myrs of an instantaneous burst the ionizing flux decreases
significantly, so that no strong emission lines would be expected.  Older
starbursts, at ages  above 7~Myr, should be detected from
their powerful FIR emission and large amounts of red supergiant stars. We
have shown indeed in Mas-Hesse \& Kunth (1991a) that the FIR luminosity
decreases with time much slower than the ionizing emission, since it 
originates mainly from the absorption of UV photons longward of the Lyman
limit. This seems to be the case of many IRAS galaxies with very weak
or absent emission lines.  NGC~4945 is a good example of a galaxy in this
post-burst phase. Koornneef (1993) has shown that mass loss from massive red
stars and supernovae from red supergiants progenitors dominate its
energetics, and has proposed the name Post-burst Infrared Galaxy for objects
in this particular evolutionary stage. We would expect a relatively high
number of supernova explosions in these galaxies hence with non-thermal
radio emission and spectral indices close to -1.0. Note that an extended
star formation process could not produce in any case such a high FIR\
output together with the weak emission lines found in these galaxies, since
ionizing photons would be produced continuously at a very high rate.

Derived ages are in any case model dependent. The
parameterization of {$W(H\beta)$} as a function of time strongly depends on the
number of ionizing photons predicted by stellar atmosphere models as well
as on the assumed number of photons directly absorbed by dust.
Moreover, if the IMF upper mass limit is below $120$~{M$_\odot$}  the age
determination would become rather imprecise during the first million years,
as shown by Leitherer \& Heckman (1995) for starbursts with upper mass
limits of only 30~{M$_\odot$}. Therefore more than the
absolute ages we can derive 
the evolutionary state of the different objects inter-compared with each
others.

Our models however are weakly sensitive to previous episo\-des of star
formation but rather to the most massive stars formed, hence the ones
younger than 10~Myr. Previous bursts of star formation would not be
detected by our method. Nevertheless, the remnants of such bursts would
still provide a significant contribution to the optical continuum. We
believe that this is the case in the few galaxies in which a significant
excess of optical emission has been detected (NGC~4670, Mrk~710, Haro~2,
NGC~4214, Tol~3 and IIZw~70). However the fact that in the majority of
these galaxies the observed {$WR/(WR+O)$} ratio is rather large rejects the
possibility that massive star formation has proceeded at a continuous rate
during such a large period of time.

\subsection {Initial Mass Function}

To isolate the effects of the IMF slope we have to select observational
parameters weakly affected by extinction and well behaved with respect to
age (i.e., monotonous functions of time). Ideal parameters are line
equivalent widths, like W(H$\beta$) and {$W(\ion{Si}{iv})$} and 
{$W(\ion{C}{iv})$}. {$W(H\beta)$} decreases
almost continuously with time and is weakly dependent on the IMF slope
(Mas-Hesse \& Kunth 1991a). On the other hand, the 
{$W(\ion{Si}{iv})\-/W(\ion{C}{iv})$} ratio increases
with time and depends on the IMF slope, since it is a good tracer of the
relative number of O3-O8 stars versus B0-B3 ones (see Fig.~1 of Mas-Hesse
\& Kunth 1991a). Therefore, combining both parameters it should be possible to
constrain simultaneously  the age and the IMF slope, as we have shown
in previous sections. Note that we are only considering
 the upper section of the IMF ($M~>~10~${M$_\odot$} ), since our method is
 insensitive to the relative amount of low mass stars (see
Mas-Hesse \& Kunth 1991a for a discussion on the effects of low mass stars).

We find that the IMF slope is constrained to the narrow range 1 to 3, with
a mean value between Salpeter one ($\alpha = 2.35$), and Scalo (1986)
($\alpha = 2.85$ for the upper section of the IMF).  Despite some scatter
in the derived values of the IMF slope, as shown in Figs.~\ref{whws}, we
find no correlation between the precise value of the slope and other
properties of the star formation episodes. We report however that objects
showing flatter IMFs are always small, compact star--forming regions, like
NGC~588, Mrk~36 and IIZw~70.  On the other hand, four of the galaxies
(NGC~4214, NGC~4670, NGC~5253 and Haro~2) with optical continuum more
clearly dominated by a previous generation of stars show rather steep IMFs
with $\alpha$ values close to 3. The previous bursts of star formation
known to have taken place in these galaxies could have modified the
properties of the interstellar clouds hampering the formation of lower mass
stars. It is interesting to note that apart {$W(H\beta)$} and
 {$W(\ion{Si}{iv})/W(\ion{C}{iv})$}, the rest of the
parameters could have been rather well fitted assuming a standard value of
the IMF slope (see for instance Rosa \& Benvenuti 1994).

Another important result from this work is that we find no correlation
between the metallicity and the slope of the IMF. The trend toward flatter
IMFs at low metallicities proposed in the past (Melnick et al. 1985;
Viallefond 1986) is therefore not confirmed by our results. This trend was
proposed to explain the apparent correlation toward higher values of the
effective temperature and {$W(H\beta)$} values at low metallicities.  Cervi\~no \&
Mas-Hesse (1994) showed that the effects on the evolution of massive stars
of decreasing the metallicity could account for this observational
trend. We have plotted in Fig.~\ref{whte} the predicted {$W(H\beta)$} and effective
temperature vs. the O/H abundance at different evolutionary stages of the
cluster, assuming nearly instantaneous star formation episodes. This figure
shows a clear trend toward higher {$W(H\beta)$} and {$T_{eff}$} values the lower the
metallicity, even with identical IMFs. We can see that at 2.75~Myr (just
before the first WR stars have been produced), {$W(H\beta)$} is higher at Z~=~{Z$_\odot$}
/20 than at Z~=~2{Z$_\odot$} by almost an order of magnitude. This effect is
even stronger at older ages.

{$W(H\beta)$} values predicted by evolutionary synthesis models during the
first 2~Myr can be as high as $\sim$500~\AA, with a fast decrease
afterwards. Nevertheless, the observational values of
{$W(H\beta)$} seldom exceed  350~\AA\ (see for example the catalogue of 
\ion{H}{ii} galaxies
by Terlevich et al. 1991). If the IMF is complete up to around 120~{M$_\odot$},
this lack of high {$W(H\beta)$} objects would indicate that almost no starburst
younger than 2~Myr has been observed. By an obvious statistical effect, we
would expect to find emission-line galaxies covering more or less
uniformly the age range between 0 and 6~Myr (when the emission lines fade),
so that this discrepancy reflects a weak point of evolutionary synthesis
models. This effect can not be attributed to flatter than normal IMFs,
since the IMF slopes in these regions are similar to that of the solar
neighborhood. The possible contamination of the optical continuum by an
underlying stellar population can neither account for this effect. We have
seen that the emission associated to the present burst generally dominates
even at optical wavelengths and that the underlying continuum is generally
negligible.  Moreover, we have been able to correct {$W(H\beta)$} from the
contamination by underlying stars. The discrepancy could be minimized
if the upper mass limit is lowered  to around 60~{M$_\odot$}. In such a case the
maximum values of {$W(H\beta)$} would remain below around 350~\AA\ at any time, in
accordance with  observational results (Mas-Hesse \& Kunth 1991a).
Nevertheless, stars with initial masses in the range
80-100~{M$_\odot$} (but not larger!) have been found in the Magellanic Clouds
(Massey et al. 1995; Heydari-Malayeri 1996), so that there are no
good reasons to truncate the IMF at around 60~{M$_\odot$} (Scalo 1986, 1990). 
Bernasconi \& Maeder (1996) have proposed recently that, since accretion
leading to the formation of massive stars should take around 2--2.5~Myr,
stars with initial masses above 40~{M$_\odot$} would emerge from its parental
cloud with a substantial fraction of its central hydrogen content already
burned. As a result, the formal Main Sequence lifetime would be reduced
and, at the time the star becomes visible, it would have already evolved to
lower effective temperatures. We postulate that this effect could be at the
origin of this discrepancy, explaining why no starbursts younger than 2~Myr
have been identified. Nevertheless, to confirm this hypothesis new models
have to be built taken into account the accretion time of the most massive
stars.  

Another possible explanation of this discrepancy could be an overestimation
of the Lyman continuum photons predicted by the models. First, the fraction
of photons that do not contribute to the ionisation process could have been
underestimated. We have assumed through this work that a fixed fraction of
30\% of the photons are directly absorbed by dust. This value corresponds
to the average fraction in Galactic \ion{H}{ii} regions, as we have discussed
above. Nevertheless, the compilation by Smith et al. (1978) shows that this
parameter covers a range between 80\% and 20\% in our Galaxy. It is
difficult to extrapolate which should be the correct value when analyzing
the integrated emission from large star--forming regions in external
galaxies, but we want to note that changing the average fraction to 60\%,
instead of 30\%, would reconcile the observations with the model
predictions.  A second possibility would be the presence of systematic
errors in stellar atmosphere models.  As discussed by Scalo (1990)
different computations can show a spread of up to an order of magnitude in
the number of ionizing photons emitted even at masses as small as
30~{M$_\odot$}. Nevertheless, Schaerer et al. (1995) have analyzed the effects of
combining stellar structure and atmosphere models on the ionizing flux,
treating consistently the entire mass loss from the center out to the
stellar winds. They find that while the shape of the ionizing continuum can
be strongly modified with respect to plane-parallel models, the total
number of ionizing photons is not significantly modified.

\begin{figure} 
\begin{center}\mbox{\epsfxsize=7cm \epsfbox{ms8377.f7}}\end{center}
\caption[]{
Predicted vs. observed {$W(\ion{Si}{iv})/W(\ion{C}{iv})$} values. Symbols as in Fig.~\ref{rwhb}. 
}
\label{rwswc}
\end{figure}

\begin{figure} 
\begin{center}\mbox{\epsfxsize=7cm \epsfbox{ms8377.f8}}\end{center}
\caption[]{
Predicted vs. observed {$L(H\beta)$} values in units of 
10$^{38}$ erg/s. Symbols as in Fig.~\ref{rwhb}.
}
\label{rhb}
\end{figure}

\begin{figure} 
\begin{center}\mbox{\epsfxsize=8cm \epsfbox{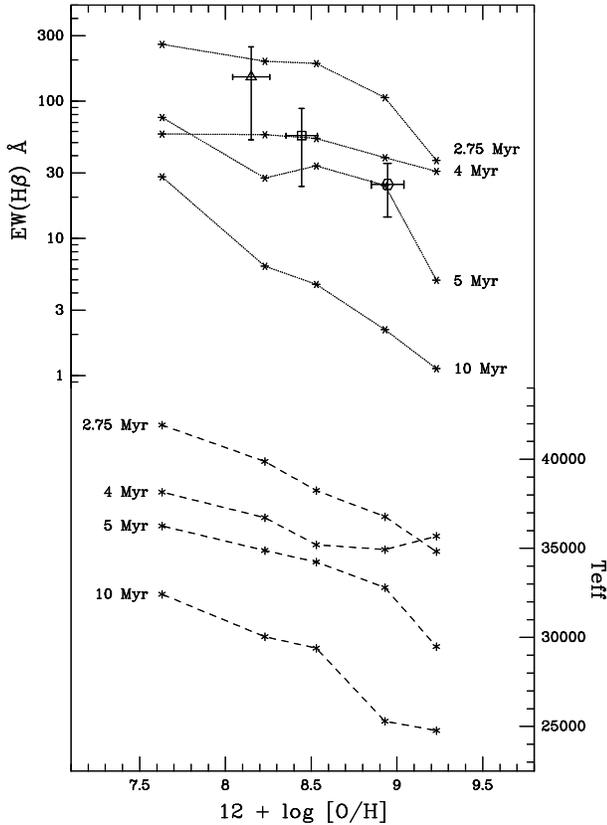}}\end{center}
\caption[]{
Predicted {$W(H\beta)$} and {$T_{eff}$} as a function of
metallicity for different ages. There is a clear tendency to smaller
values of both parameters the higher the metallicity.
}
\label{whte}
\end{figure}

If massive stars formation proceeds as episodic instantaneous bursts, the
formation of low mass stars could be seriously hampered.  The time needed
for the protostellar collapse of a low mass cloud is at least several
million years, in any case longer than the time required for the formation
of the most massive stars. Therefore, if we assume that the onset of
massive stars inhibits further star formation, low mass could be very
deficient in these regions. Nevertheless, Zinnecker et al. (1996) have
found from X-ray and IR imaging that OB associations contain stars with
masses below 1~{M$_\odot$}, and that 30~Dor-like associations may be surrounded
by low mass stars. Low mass stars seem therefore to have formed prior to
the massive stars, as postulated by Tenorio-Tagle (1994). Under
this scenario the whole population of low mass stars would form at the very
beginning of the collapse. Shocks induced in the gas by the movement of
these stars would trigger the collapse of more massive stars. These massive
stars would finally heat the gas, inhibiting further star formation. In
this way, the whole IMF would be filled with stars.

\subsection {Wolf-Rayet stars population}

We have plotted in Figs.~\ref{wr}a,b the predicted 
{$L(WR)/L(H\beta)$} ratios and {$W(WR_{bump})$}\
for different metallicities, two limiting values of the IMF slope and both
star formation regimes.  We have included in Figs.~\ref{wr}a,b the mean
values of the {$L(WR)\-/L(H\beta)$} ratios and {$W(WR_{bump})$} of a more complete sample of galaxies
compiled from Kunth \& Joubert (1985), Kunth \& Schild (1986) and Vacca \&
Conti (1992) (see also the comparisons made by Maeder
\& Meynet 1994 and Meynet 1995).  Points have been located at their mean
ages, as estimated by the {$W(H\beta)$} values.  Error bars indicate the range of
ages covered. Nevertheless, these age estimates are only approximate, since
no correction at all has been applied to the observational values,
generally taken through narrow slits.

\begin{figure}[t] 
\begin{center}\mbox{\epsfxsize=8cm \epsfbox{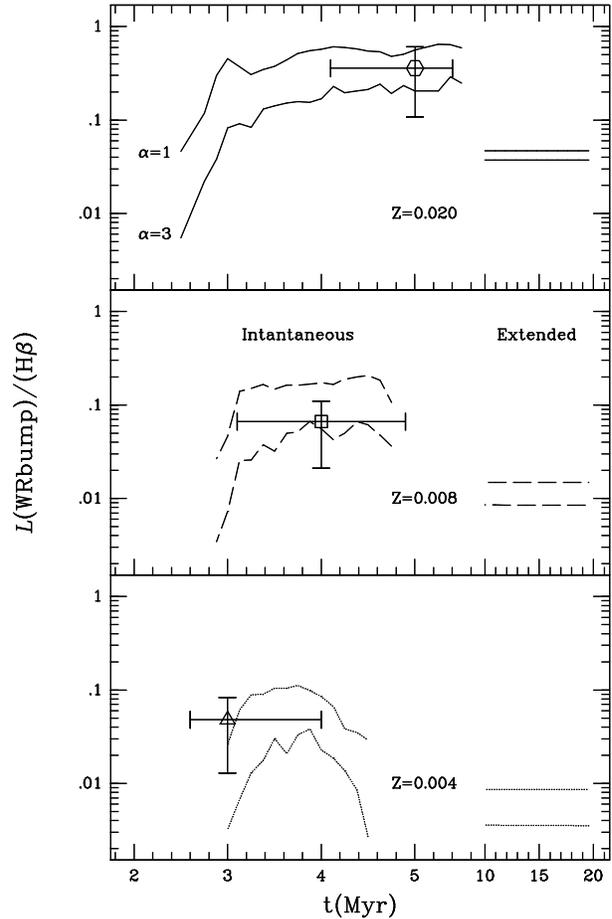}}\end{center}
\caption[]{
a) {$L(WR)/L(H\beta)$} ratio and b) $W(WR_{bump})$  predicted by
the models. We have plotted here the predictions for the limiting values of
the IMF slope considered and for both instantaneous and extended star
formation regimes, this latter one only for ages above 10~Myr. Mean values
from an extended sample of objects (see text) have been included. Note that
they are generally located within the predicted bands. 
}
\label{wr}
\end{figure}

\addtocounter{figure}{-1}

\begin{figure}[t] 
\begin{center}\mbox{\epsfxsize=8cm \epsfbox{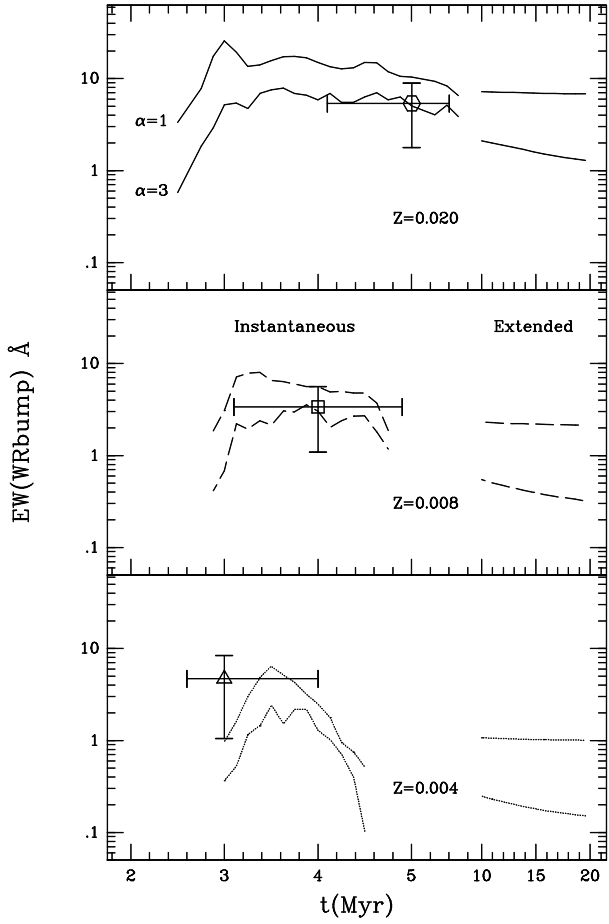}}\end{center}
\caption[]{\em Continuation }
\end{figure}

It can be seen that our results can be extended to this more complete
sample. The average {$W(H\beta)$} and {$W(WR_{bump})$} values fall fairly well within the
limiting predicted curves. It is also remarkable that the derived range of
ages coincide rather well with the ages at which WR stars are expected.
Although the lower limits of the error bars would be in some few cases
compatible with extended star formation regimes, the average ratios suggest
again that the star formation episodes should have taken place almost
instantaneously.  The fact that also in this more extended sample the
agreement with the {$W(WR_{bump})$} values is satisfactory indicates that in these
galaxies the burst emission largely dominates in general over the older
underlying stellar population.

Schaerer (1996) has evaluated the effect of the evolution of the WR stars
population on the \ion{He}{ii} narrow emission line at 4686~\AA , by combining
model atmospheres accounting for stellar winds with evolutionary tracks.
He concludes that for metallicities in the range Z = {Z$_\odot$} /5 -- {Z$_\odot$} a
strong nebular \ion{He}{ii} emission line should  originate in early WR phases
 when WC stars begin to appear. The \ion{He}{ii} emission line is
indeed detected in few objects with very young ages,
below 3~Myr, and therefore starting to produce WR stars (NGC~2363 and
Mrk~36), in good agreement with the scenario proposed by Schaerer.

\subsection {Supernovae and radio emission}

The relatively large number of massive stars in these small regions
originate frequent supernova explosions after the first 3~Myr, when the
most massive stars (initial mass around 120~{M$_\odot$} ) end their lifetime. The
evolution of the supernova rate depends strongly on the IMF slope, as
discussed in Cervi\~no \& Mas-Hesse (1994).  For slopes between $\alpha =
2$ and 3 the supernova rate varies only weakly with time. Around
$10^{-9}$ yr$^{-1}$ {M$_\odot$}$^{-1}$ supernovae explosions are expected with an
increasing supernovae rate the higher the metallicity.  Considering the
mass involved in the star--forming episodes, we expect supernova rates in
the range $10^{-7}$ (NGC~588) to $10^{-3}$ yr$^{-1}$ (Haro~2). Typical
values for compact galaxies are around 1 supernova each 1000 years and
remains almost constant after the first 3~Myr, up to around 50~Myr, when
stars of initial mass around 6~{M$_\odot$} end their lifetime.  Therefore,
thousands of supernova explosions must have taken place within the relatively
small star--forming regions of the most evolved objects we have studied.

The large number of supernova remnants (SNR) must become strong sources of
non-thermal radio emission. Depending on the evolutionary state of the star
formation episode the total radio emission should be dominated by nebular
thermal free-free emission, with a flat radio spectral index ({$\alpha_{radio}$} $\sim
-0.1$) or by the non-thermal contribution from SNR ({$\alpha_{radio}$} $\sim -0.9$)
(see Mas-Hesse 1992 for a more detailed discussion of radio emission from
starbursts).

\begin{figure} 
\begin{center}\mbox{\epsfxsize=8cm \epsfbox{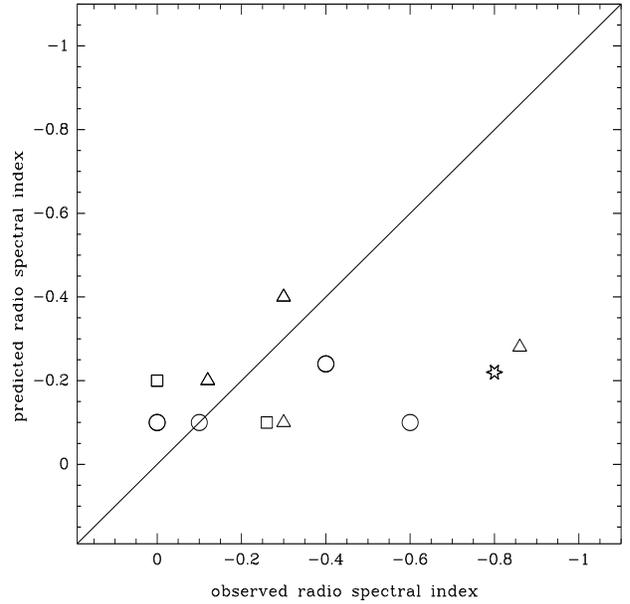}}\end{center}
\caption[]{
Predicted vs. observed radio spectral indices around 6~cm. Symbols as in
Fig.~\ref{rwhb}.  
}
\label{ralfa}
\end{figure}

\begin{figure} 
\begin{center}\mbox{\epsfxsize=8cm \epsfbox{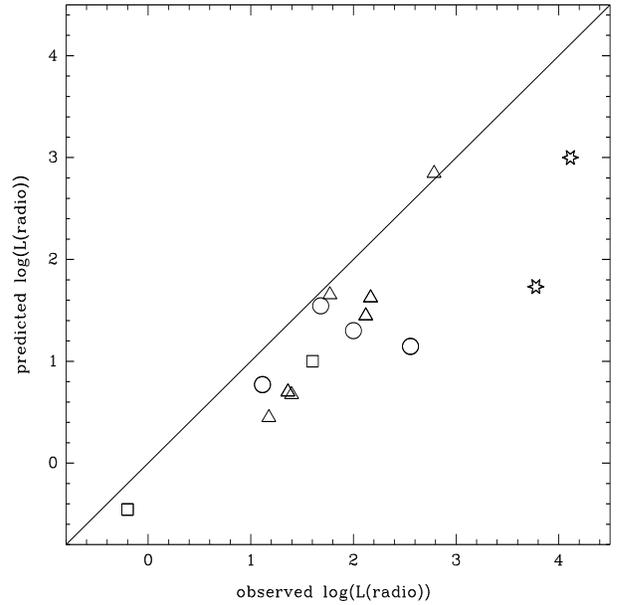}}\end{center}
\caption[]{
Predicted vs. observed radio luminosities at 6~cm in units of 10$^5$ Jy kpc$^2$. 
Symbols as in Fig.~\ref{rwhb}. }
\label{lalfa}
\end{figure}

We have plotted in Figs.~\ref{ralfa} and \ref{lalfa} the predicted vs. the
observed radio spectral indices and radio luminosities at 6 cm,
respectively. It is clear that our models significantly underestimate the
observed radio luminosities.  The predicted emission is underestimated by
factors between 2 (NGC~595, NGC~2363 and Mrk~59) and around 5 for the
majority of BCGs (IZw~18, Mrk~36, IZw~36, NGC~4670, NGC~5471 and
IIZw70). In all these objects the thermal contribution is dominating, since
they are younger than 4~Myr and have had no time for a significant number
of supernova explosions. The radio spectral indices are indeed above
{$\alpha_{radio}$} ~=~-0.3 (except for Mrk~59), as predicted by the models.  The radio
emission in IZw~18, an object having experienced apparently a sequence of
bursts during the last 13~Myrs, is also purely thermal, showing that the
contribution of non-thermal sources becomes negligible when a large number
of ionizing stars are present.

The thermal radio emission is a known function of the number of ionizing
photons emitted. Since our models are matched to reproduce  the
intensity of the Balmer lines, which are also a linear function of the
number of ionizing photons, there should be a good agreement between the
observed and predicted thermal radio emissions, but this is not the
case. The discrepancy between the observed Balmer lines emission and the
thermal radio luminosity is a well-known problem (Israel \& Kennicutt 1980;
Lequeux et al. 1981; Caplan \& Deharveng 1986; Skillman \& Israel
1988). The most accepted explanation is that the effective reddening in the
radio range is smaller than in the optical, due essentially to geometrical
effects and a patchy distribution of dust. Nevertheless, since the massive
stellar population we  derive reproduces well the de-reddened Balmer
line fluxes (within less than a factor 1.6 in average), the intrinsic
total number of ionizing photons injected to the nebular gas should be well
constrained. A possible explanation for this discrepancy would be the
presence of a large number of ionizing stars in small regions completely
obscured by dust in the optical, from which only the associated radio
emission could escape (and partially  also the near infrared emission lines).
 This scenario is nevertheless not consistent with our
results, since the large number of hidden stars we would need to reproduce
the observed radio emission would increase significantly the observed FIR\
luminosities contrary to what we observe (see next section).
 Inhomogeneous scattering, depending on
the specific geometrical distribution of dust in each galaxy and with
different efficiencies at radio and optical ranges, might provide the
additional contribution to the observed radio emission 
(Caplan \& Deharveng, 1986). If this explanation is valid, a word of caution
should be given against deriving intrinsic extinctions from the observed
ratio between radio and Balmer lines emission, since the reddening can be
significantly overestimated.

For larger galaxies the underestimation of the radio emission is
even more severe, between a factor 10 for Haro~2 and two orders of
magnitude for Mrk~710. These discrepancies can be partly attributed to
aperture mismatch, especially important in these large galaxies, as well as
to the expected contribution of supernova remnants from previous episodes
of star formation. As we have discussed above, supernovae events can be
active until more than 50~Myr, when the stellar contribution to the UV
continuum and to the ionisation process has become negligible. In this
sense it is interesting to note that the three galaxies for which the
discrepancy is larger (Haro~2, NGC~4214 and Mrk~710) show radio spectral
indices below {$\alpha_{radio}$} = -0.4, i.e. dominated by non-thermal sources.

\subsection {Far Infrared Emission }

The Far Infrared emission in objects dominated by intense star formation
episodes is essentially due to thermal emission by dust particles heated by
the UV stellar radiation field. This process of absorbing high energy
photons, re-emitted at low energies (typically at around 100~$\mu$m),
can be so efficient that the FIR luminosity becomes  a good
approximation to the bolometric luminosity. Since the origin
of this FIR emission is mainly the energy output of massive stars, whose
UV photons are efficiently absorbed by dust, {$L_{FIR}$} has been used in the
past to derive the number of massive stars in the young clusters, and
henceforth the star formation rate (Hunter et al. 1986).

The technique we have used to fit the unreddened synthetic SEDs to the
actually observed ones provides a direct estimation of the expected total
FIR luminosity. As we have explained in detail in Mas-Hesse \& Kunth
(1991a) we compute the total energy absorbed by dust from the
difference between the unreddened synthetic SED and the reddened SED that
best fits the observed spectrum. We take into account the energy provided
by Lyman $\alpha$ photons which are absorbed by dust and a fixed fraction
of 30\% of the Lyman continuum photons which are absorbed by dust and do
not contribute to ionisation. 

Calzetti et al. (1995) have estimated the expected contribution to
the FIR emission of a large sample of star-forming galaxies by assuming a
single extinction law (Calzetti et al. 1994) and a
color excess derived from the Balmer decrement, decreased by 
0.5 mag. While this approach might be useful to derive average properties of a
large sample, our procedure derives first both the shape and the strength
of the extinction law affecting the continuum, and uses  these results
to compute the expected FIR luminosities.  These authors separate the
FIR emission coming from cool ($T_{dust}$ around 20 K) and warm
($T_{dust}$ around 40~K) dust, as also proposed by Buat \& Deharveng
(1988). In our
sample star formation processes are so intense that the FIR emission is
dominated by the warm dust. The mean dust temperature is around 40~K, as
shown in Table~\ref{rfir}, so that a two-components deconvolution is not
needed. Furthermore, the emission in the IRAS bands below 60~$\mu$m is
generally very low and has not been detected, so that the extrapolation to
the whole FIR emission based on the 60 and 100~$\mu$m fluxes can be taken
as a good estimation of {$L_{FIR}$}.

\begin{figure} 
\begin{center}\mbox{\epsfxsize=8cm \epsfbox{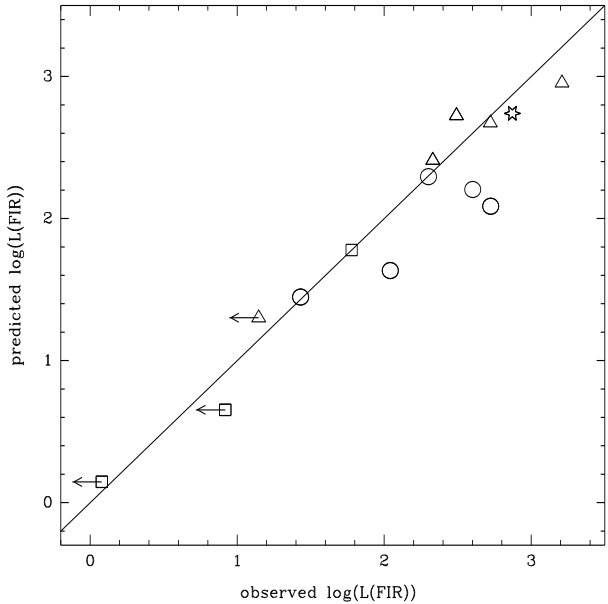}}\end{center}
\caption[]{
Predicted vs. observed FIR luminosities in units of 10$^{40}$ erg/s. 
Symbols as in Fig.~\ref{rwhb}.}
\label{lfir}
\end{figure}

We have plotted in Fig.~\ref{lfir} the predicted vs. observed FIR\
luminosities.  It can be seen that both values are in general in good
agreement.  {$L_{FIR}$} is underestimated only in the case of spatially extended
galaxies since while the predictions have been normalized to the energy
emitted within the $10\arcsec\times 20\arcsec$ IUE aperture, IRAS fluxes were integrated over
the whole ga\-laxy. The discrepancy is highest for NGC~4214, in which there
are known sites of massive star formation outside the IUE aperture
(Ma\'{\i}z-Apell\'aniz et
al. 1998). As we have discussed in the previous section, the good agreement
between predictions and observations, especially for spatially compact
objects, indicates that we are sampling the whole population of massive
stars, even at UV wavelengths (the measured UV continuum is used for
de-normalization), and that the contribution from stars hidden within dust
clouds is negligible. The only exceptions are NGC~595 and NGC~3256, a very
dusty galaxy, in which only a small fraction of massive stars contribute to
the UV continuum while the majority of the newly formed stars are
apparently completely obscured by dust. Calzetti et al. (1995) have also
found that an average of 30\% of massive stars might be embedded in dusty
regions within dusty galaxies, where {$E(B-V)$} $>$ 0.25.

\subsection {Extinction}

Our sample is biased towards low reddening galaxies with strong UV
continuum with the exception of NGC~3256, selected by its strong FIR
emission and showing by far the strongest extinction. Nevertheless, as we
have discussed above, even a small amount of dust can significantly modify
the strength and shape of the UV continua.

\begin{figure} 
\begin{center}\mbox{\epsfxsize=8cm \epsfbox{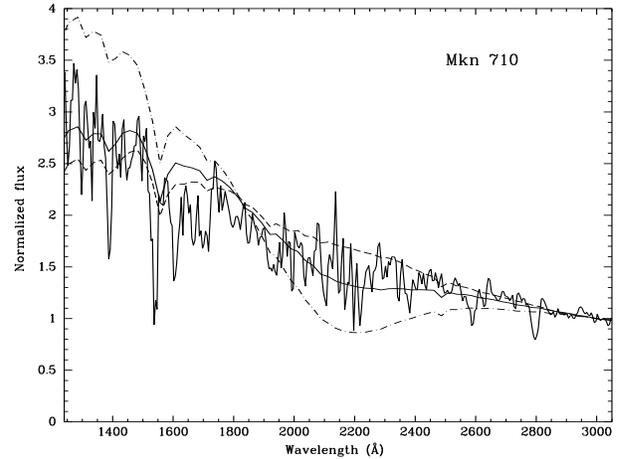}}\end{center}
\caption[]{
Synthetic spectra reddened assuming three different extinction laws on top
of the observed UV continuum of Mrk~710. Solid: best fitting law (LMC);
dot-dashed: Galactic law; dashed: SMC law.
}
\label{red}
\end{figure}

From our analysis it becomes evident that the dust absorption feature at
around 2175~\AA\ is very weak in these galaxies or even
inexistent. Relatively weak absorption bumps in starburst spectra have
already been noted by several authors (Blitz et al. 1981; Lequeux et
al. 1981; Rosa et al. 1984; Kinney et al.  1993). The weakness of this
absorption feature induced us to include both the LMC and SMC extinction
laws in our fitting procedure. It is important to note that these laws have
a rather different shape in the UV than the standard Galactic one, apart
from  a weaker (LMC) or inexistent (SMC) absorption
feature. Therefore, even when only the 1200-1950 IUE range is available
it is possible to discriminate which of the laws is
applicable. The method we have followed is very sensitive to
both the shape of the law and its strength, discriminating between
different laws for color excesses above {$E(B-V)$} =~0.05, with
{$E(B-V)$} being determined at a precision of $\delta(E(B-V)) = 0.02$. 
An example is shown in Fig.~\ref{red}, where we have plotted the best fit
obtained assuming three different extinction laws.

In almost all cases interstellar extinction closely follows either an LMC
or SMC-like law. Since there is a tendency to have weaker absorption
features at 2175~\AA\ at lower metallicities in nearby galaxies (Galaxy -
LMC - SMC) we would expect a similar trend in our sample. This is not the
case since even near-solar metallicity galaxies like Mrk~710 and NGC~3256
are best fitted assuming an LMC extinction law and not a Galactic one. The
only object in our sample clearly affected by a Galactic-like extinction is
NGC~595.  Calzetti et al. (1994) have derived an universal UV to optical
extinction law from the SED of 39 starburst and blue compact galaxies. This
law and the Galactic one show similar optical-far UV slopes, but the
Calzetti et al. (1994) one is essentially featureless. We stress here that
the observed UV to optical SEDs can be very well reproduced by reddening
the corresponding synthetic spectra with one of three extinction laws
(Galactic, LMC and SMC) with no need to invoke an additional universal law.

The weakness of the absorption bump at 2175~\AA\ seems to be related to the
intense star-formation processes taking place in these objects, rather than
to the metallicity. The strong flux of high energy photons around the young
massive clusters and the relatively high temperature that dust particles
can reach in these environments affect most probably the size and
composition of dust grains. Comparing the shape of the Galactic, LMC and
SMC extinction laws with the absorption cross sections by amorphous
silicates and graphite grains given by Mezger, Mathis \& Panagia (1982) we
conclude that graphite grains are deficient in starburst regions.  An
interstellar medium dominated by amorphous silicates would explain
naturally both the weakness of the 2175~\AA\ bump and the steepness of the
extinction law.  Puget \& L\'eger  (1989) attribute the UV absorption bump
to carbonaceous very small particles containing polycyclic aromatic
molecules. They describe how a combination of photothermodissociation,
double ionization and Coulomb explosion processes induced by an intense and
hard UV radiation field would decrease significantly the abundance of these
particles. We suspect that these processes do not depend on the metallicity
of the environment, but on the intensity of the UV emission, explaining the
lack of correlation between the shape of the extinction law and the nebular
metal abundance.

\begin{figure} 
\begin{center}\mbox{\epsfxsize=8cm \epsfbox{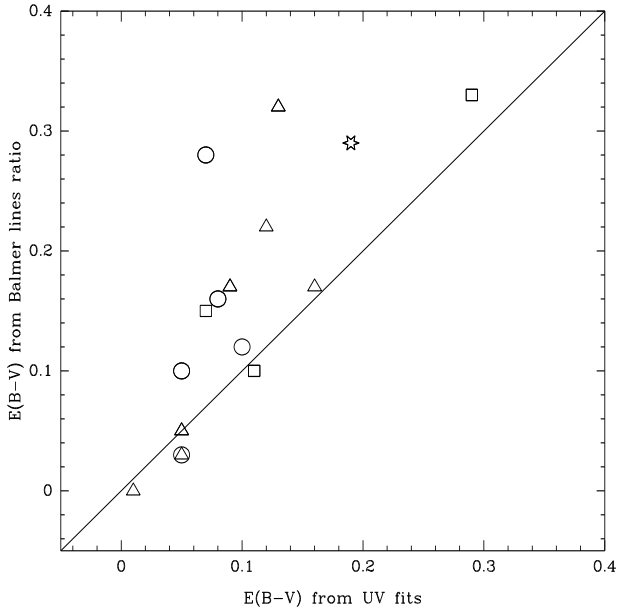}}\end{center}
\caption[]{
{$E(B-V)$} values derived from Balmer lines ratios vs. values obtained from our
UV continuum fitting procedure. Symbols as in Fig.~\ref{rwhb} (NGC~3256
extreme case ({$E(B-V)$} =~0.85 vs. 0.16) has not been included in the plot for
scaling reasons). 
}
\label{uvo}
\end{figure}

We have found that for several galaxies the {$E(B-V)$} values derived from the
Balmer line ratios are systematically higher than values derived from the
UV continuum (see Table~\ref{res}). This discrepancy can be clearly
appreciated in Fig.~\ref{uvo}, where we have plotted both {$E(B-V)$}\
determinations.  A similar effect has already been noticed by Fanelli et
al. (1988), Keel (1993) and Calzetti et al. (1994) in the spectra of
starburst galaxies.

It is important to remark that his UV-optical discrepancy is related to two
very different sources of emission: stellar continuum and nebular emission
lines.  Two possible explanations for this discrepancy are the presence of
strong absorption lines of stellar origin and geometrical effects. The
usual correction for the contamination by stellar lines assumes absorption
equivalent widths of around 1-2~\AA\ and iterates until the color excesses
derived from the $H\alpha/H\beta$, $H\beta/H\gamma$ and $H\beta/H\delta$
ratios converge to the same value (McCall, Ribsky \& Shields 1985). This is
the method we have followed to derive the {$E(B-V)$} values, but it does not
solve the discrepancy. On the other hand, from the synthesis of the Balmer
lines absorption profiles, D\'{\i}az (1988) and Olofsson (1995) 
proposed that the stellar
equivalent widths should be much larger, up to around 5~\AA. We have
therefore forced the {$E(B-V)$} value to be the same as obtained from UV
continuum fits, and have derived so the corresponding equivalent widths of
the absorption lines. In some cases the method converged, yielding
consistent values of the absorption equivalent widths between 4 and 5~\AA\
for the three Balmer lines {H$\beta$}, H$\gamma$ and $H\delta$. But for the
majority of the objects the equivalent widths did not match the relative
strengths predicted by D\'{\i}az (1988).

\begin{figure} 
\begin{center}\mbox{\epsfxsize=8.5cm \epsfbox{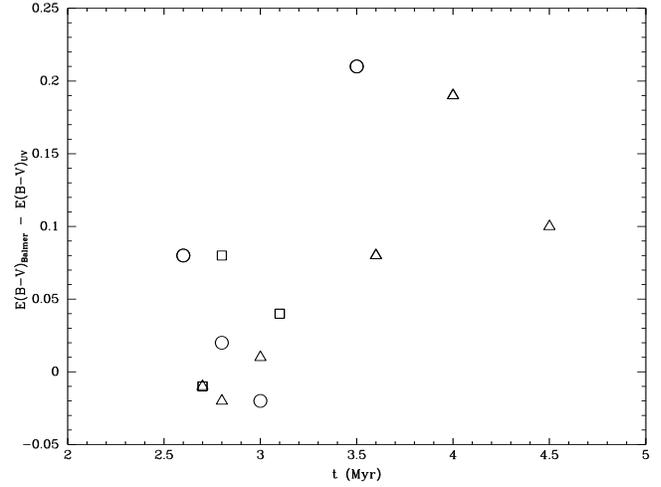}}\end{center}
\caption[]{
{$E(B-V)$} values discrepancy as a function of derived age.  Symbols as in
Fig.~\ref{rwhb}. 
}
\label{ebvt1}
\end{figure}

\begin{figure} 
\begin{center}\mbox{\epsfxsize=8.5cm \epsfbox{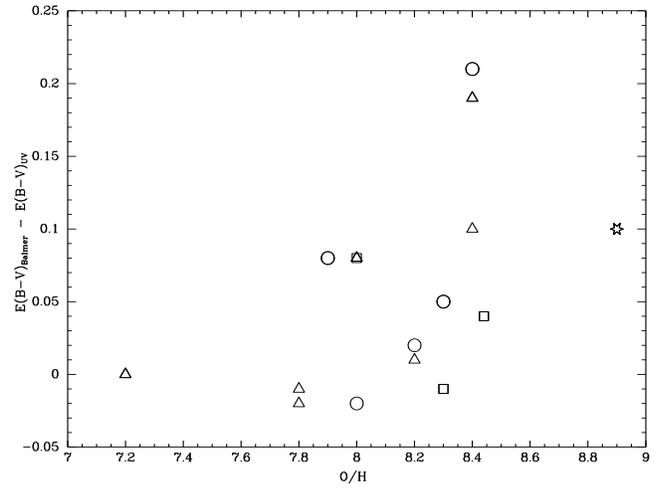}}\end{center}
\caption[]{
{$E(B-V)$} values discrepancy as a function of nebular O/H abundance.  Symbols
as in Fig.~\ref{rwhb}. 
}
\label{ebvt2}
\end{figure}

The underlying stellar population seems to be responsible for this
discrepancy in the case of Haro~2, at least partially. The method converged
well, deriving stellar equivalent widths of 5.1, 4.2 and 4.1 \AA\ for {H$\beta$},
H$\gamma$ and $H\delta$, respectively. Assuming this correction, the
derived {$E(B-V)$} coincides with the value estimated from the UV continuum
fits. Moreover, we have recently got high dispersion optical spectra of
this galaxy and have been able to measure the equivalent widths of the
stellar absorption lines (Legrand et al. 1997a), obtaining values in good
agreement with the predicted correction ($W(H\gamma) = 4.1\pm1.0$ \AA;
$W(H\delta) = 4.7\pm1.0$ \AA).  We conclude that in some galaxies, the
discrepancy between the strength of the extinction derived from the
continuum and from the Balmer decrement can be attributed to the presence
of strong stellar absorption lines, associated to an underlying population
of relatively low mass stars.

In NGC~4214, on the other hand, the required equivalent widths derived by
this method varied between 13 and 6~\AA\ for the different lines. Since
such a large range of equivalent widths is not expected for the stellar
absorption lines, we attribute the discrepancy to
geometrical effects related to the inhomogeneous distribution of
dust. While the continuum flux comes from the young clusters, the nebular
emission originates in much more extended regions well beyond the ionizing
sources. The strong ionizing flux is expected to affect the dust particles
properties in the close surroundings of the massive stars.  Stellar winds
and supernova explosions might wipe out the dust from the neighbourhood of
the massive clusters and concentrate it in filaments and dust patches still
located within the ionized region. Depending on the specific geometrical
distribution of dust and stars, the extinction could affect mainly the
nebular gas emission but only weakly the stellar continuum. 

In the specific case of NGC~4214, as we will discuss later,
Ma\'{\i}z-Apell\'aniz et al. (1998) have found that the Balmer line
emission peaks at few arcseconds away from the massive stars
cluster. Moreover, the reddening is highest in regions where Balmer lines
are emitted while it is relatively low in front of the star cluster. The
reddening in front of the cluster is consistent with that derived from
continuum fits, while the reddening derived integrating spatially over the
whole ionized region is significantly higher. These results have shown that
an inhomogeneous distribution of dust can explain indeed the different
extinctions derived from the stellar continua and the Balmer
decrement. Nevertheless, this effect is not universal and depends strongly
on the distribution of stars, gas and dust in each star--forming region. In
the second brightest knot of star formation in NGC~4214 for example, the
distributions of stars, ionized gas and dust are all centered in
essentially the same regions. Our results are in good agreement with
Gordon, Calzetti \& Witt (1997), who also conclude from the analysis of the
extinction in 30 starbursts that the stars/dust geometry should be close to
a dust-free sphere of stars surrounded by an outer star-free shell of
clumpy dust.  We want to remark that since our method yields the extinction
affecting the stellar continuum only, it might be well that the dust
particles producing the UV bump have been destroyed from the neighbourhood
of the massive stars, but they could still be present in the ionized gas
located far away from the massive clusters.

We have found a trend toward higher $E(B-V)_{Balmer} - E(B-V)_{UVcont}$
values the older the star-forming episode, as shown in
Fig.~\ref{ebvt1}. There is a well defined lower envelope, rising with age.
On the other hand, note that the trend is very weak or even inexistent if
we compare $E(B-V)_{UVcont}$ with the age of the cluster (see
Table~\ref{res}). Similarly, although no clear trend with metallicity is
found, the $E(B-V)_{Balmer} - E(B-V)_{UVcont}$ discrepancy becomes highest
for galaxies with O/H abundances above 8.4, i.e., objects whose
interstellar medium has already been polluted by previous star formation
episodes (see Fig.~\ref{ebvt2}).  We interpret these results as an evidence
for an interstellar medium enrichment with dust during the evolution od the
clusters.  This fraction of the extinction which is raising with time could
be due to newly formed dust grains formed in stellar winds and in the
supernova ejecta. In both cases dust would have been blown away from the
stellar cluster, concentrating on given regions, still within the ionized
nebula, but at a certain distance from the massive stars. In this way it
would have a stronger effect on the emission lines than on the stellar
continuum.

The discrepancy discussed hereabove has important implications on the
analysis of starburst regions. First of all, the {H$\beta$} equivalent width,
which is used as an age indicator can be affected by this differential
reddening. A difference of $\Delta(E(B-V))=0.2$ between nebular and
continuum extinction implies an underestimation in the value of {$W(H\beta)$} by a
factor of 2. Such discrepancies are not rare, as shown by Calzetti et
al. (1995). They can lead to an overestimation of the age in some
star--forming regions. A similar effect occurs when measuring the {$L(WR)/L(H\beta)$}\
ratio, since the WR bump originates in the atmosphere of these stars and
should therefore be affected by the same extinction than the stellar
continuum. Unfortunately, the only way to correct properly for this effect
requires the mapping of the distribution of stars, gas and dust, something
which is feasible only for nearby galaxies.

\section {Analysis of some individual galaxies}

We discuss in the next paragraphs the results we have derived for some
representative individual objects. 

\subsection{NGC~5471} 

NGC~5471 is a massive G\ion{H}{ii} region located in M101. 
Our analysis yields a scenario characterized by a young (around 3.0~Myr),
massive (1.2$\, 10^5$~{M$_\odot$} ) cluster, following a standard IMF (slope
slightly higher than the Salpeter value). Our age determination coincides
with the value derived by Rosa \& Benvenuti (1994) from HST FOS data, also
based on the comparison with evolutionary synthesis models. NGC~5471 seems
to be dominated by a well defined burst of star formation with a negligible
contribution from older underlying stars. All the
observational parameters can be reproduced by our models and only the radio
emission is underestimated, as for the majority of the objects.

The effective temperature is  high and hints toward the presence of
stars more massive than at least 40~{M$_\odot$}.  Furthermore, the known
existence of WR stars in this object of relatively low metallicity (O/H
around 8.0) indicates that the IMF upper mass limit has to be very close to
at least 100~{M$_\odot$}, since only stars more massive than around 80~{M$_\odot$}\
can become WR at this metallicity. 

Reddening in NGC~5471 is moderate, with a derived color excess {$E(B-V)$} =
0.07.  At this low reddening it becomes difficult to disentangle between
the various extinction laws, but the Galactic one provides the best
fit. This color excess derived from fitting the continuum is significantly
smaller than the value derived from the Balmer line ratio by Rosa \&
Benvenuti (1994), {$E(B-V)$} = 0.15.  The good agreement between the predicted
and the observed UV continua, as well as the FIR luminosities, let us
conclude that the continuum cannot be reddened by more than {$E(B-V)$} =
0.07. This conclusion is reinforced by the good agreement of our
1200-7000~\AA\ synthetic spectrum with the complete HST FOS spectrum
of Rosa \& Benvenuti (1994). As
shown in Fig.~\ref{spec1}, the fit is indeed rather good in the whole range,
except for an overestimation of the continuum between 3700 and 4900~\AA.
The small region observed with FOS seems  to be representative of
the whole cluster, as observed through the large IUE aperture.

A broad {H$\alpha$} emission line (FWHM~$\approx$~2000 km s$^{-1}$)
was detected in this region by Casta\~neda et al. (1990) and was
confirmed by Mas-Hesse et al. (1994).  The origin of these broad line
components is not yet well understood. One of the possible hypothesis
relates such broad components with supernova activity.  At the age derived
for NGC~5471, close to 3~Myr, no supernovae should have yet taken place,
since even stars with initial masses up to 120~{M$_\odot$} need longer than
3.5~Myr to end their lives as supernovae.  Therefore, our results provide a
hint against this origin for the broad lines, and point rather to the
effect of strong stellar winds, as proposed by Tenorio-Tagle et al. (1997)
for other starbursts. 

\subsection{NGC~4214}

NGC~4214 is a rather large irregular galaxy. It shows a central bar with
several star--forming knots. The presence of incipient spiral arms  led
to its classification as SBmIII (Sandage \& Bedke 1985). It has been
scanned spectroscopically in the optical by Ma\'{\i}z-Apell\'aniz
 et al. (1998), who
have produced maps of the physical properties (extinction, density,
excitation,...) and have also found line splitting and broad components at
several locations. The overall view resulting from these observations is
that several short lived star formation episodes have taken place over the
bar connecting the two spiral arms of this galaxy, with a significant
 age spread (few Myr). Spectra and images of the brightest star--forming
region have also been obtained with the Hubble Space Telescope by Leitherer
et al. (1996), showing the presence of a bright, compact starburst knot
surrounded by more than 200 fainter point-like sources. This compact
region has been observed with IUE, including  within its aperture a
fainter knot located at $8\arcsec$ to the East (Huchra et al. 1983).

Our best fitting models yield an age of around 3 - 3.5~Myr for this central
star formation knot. The large number of WR stars detected indicates that
this burst has been essentially instantaneous. Both the {$L(WR)/L(H\beta)$} ratio and the
equivalent width of the WR bump constrain the age of the burst to
between 3 and 5 Myr old. A broad WR feature attributed to \ion{C}{iv}$\lambda$5808
emission was detected in the two star-forming knots within the IUE aperture
(Mas-Hesse \& Kunth 1991b, Sargent \& Filippenko 1991). The ratio between
this line and the bump at around 4686~\AA\ has been synthesized by Meynet
(1995). The ratio we have measured, $L(WR_{4686})/L(WR_{5808})$ = 0.5, is
also consistent with the age constraints we have derived. It is interesting
to note that the high value of this ratio can only be reproduced assuming
enhanced mass-loss rates in the WR atmosphere (Meynet 1995).  At the
metallicity of NGC~4214, the peak in the WC/WR ratio is reached between 3
and 3.5 Myr. {$W(\ion{Si}{iv})/W(\ion{C}{iv})$} and {$W(H\beta)$} are simultaneous consistent with an age for
the burst around 3~Myr. They furthermore constrain the IMF slope to a
rather flat value close to $\alpha = 3$. This short discussion serves as an
example of how the use of different parameters (WR bump, {$W(H\beta)$},
 {$W(\ion{Si}{iv})/W(\ion{C}{iv})$} )
allows to constrain uniquely the properties of the star formation
episode. The fact that all they predict the same range of ages provides
a strong reliability on the consistency of the method. 

We want to remind here that the optical continuum is seve\-rely contaminated
by an underlying population of older stars (see Fig.~\ref{spec1}). Both
{$W(H\beta)$} and {$W(WR_{bump})$} had to be corrected from this underlying continuum before
comparison with the predictions of synthesis models. Otherwise, the results
would not have been reliable. Indeed, without removal of the contamination
it would have been impossible to fit all the parameters
simultaneously. Moreover, in the case of extended regions like this one,
special care has to be given to integrate properly the continuum from the
young cluster and the emission lines originating in the whole extension of
the gas ionized by the cluster.  In this case, both the {$L(WR)/L(H\beta)$} and {$W(H\beta)$}\
corrected values we have used in the fitting have been taken from
Ma\'{\i}z-Apell\'aniz et al.  (1998) and were obtained, integrating the
whole {H$\beta$} emission associated to the star cluster, correcting from
differential extinction and removing the optical continuum contribution
from older stars.  These authors have
shown, for example, that the {$L(WR)/L(H\beta)$} value measured over an aperture of
$2\arcsec\times 2\arcsec$ could be up to 25 times larger than the intrinsic
value. This effect led Sargent \& Filippenko (1991) to conclude erroneously
that the WR population in this galaxy was abnormally high. Similarly, the
value of the {$L(WR)/L(H\beta)$} ratio quoted by Meynet (1995) was obtained through a
small aperture and is severly overestimated: it can't indeed be reproduced
by any synthesis model.

The IMF slope we have derived coincides with the optimum value obtained by
Leitherer et al. (1996) by fitting the profiles of the \ion{Si}{iv} and \ion{C}{iv} stellar
absorption lines oberved with HST. Although they derive a slightly older
age (around 4~Myr), this good agreement provides an additional support for
the validity of the assumptions we have made for using the 
{$W(\ion{Si}{iv})/W(\ion{C}{iv})$} ratio, as
we have explained above. 

NGC~4214 is one of the objects  showing the largest
discrepancy between the reddening derived from our continuum synthesis
procedure and the extinction measured through the Balmer lines ratios
({$E(B-V)$} = 0.07 vs. 0.28, respectively). As we have discussed above,
Ma\'{\i}z-Apell\'aniz et al.
 (1998) have mapped the distribution of stars, ionized gas
and dust in this object, showing that the dust is concentrated at a certain
distance from the stellar cluster, but still within the ionized nebula. As
a result, the UV continuum comes out without being  significantly affected by
extinction, while the emission lines are strongly affected by dust. The
quite large number of massive stars formed have apparently
wiped out the dust particles from their surroundings, leaving free paths
through which the stellar continuum can emerge. We believe that this
scenario could also be valid for other large star--forming regions affected
by strong stellar winds.

We  finally want to comment that the observed FIR luminosity is
underestimated by the models, as expected since we know that there are several
star--forming knots outside the aperture we have used for normalization,
that still contribute to the IRAS measurements. 

\subsection{Mrk~59} 

Mrk~59 is a giant \ion{H}{ii} region located to the southern extreme of the
irregular galaxy NGC~4861. Its UV spectrum shows a very steep continuum,
indicating the absence of significant amounts of dust. The optical emission
lines are also unaffected by reddening. The region is dominated by a strong
burst having transformed around 10$^6$ {M$_\odot$} of gas into stars some 3~Myr
ago. The IMF seems to be rather steep, although its determination is rather
inaccurate for this object. There is an important population of WR stars,
as derived from the {$L(WR)/L(H\beta)$} ratio, which at the metallicity of this galaxy
excludes the possibility of an extended star formation episode and
constrains the ages to the range 3.0 - 4.5 Myr.

The global absolute magnitudes as {$L_{FIR}$} and {$L(H\beta)$} are well reproduced by
the models (within 15\%). It is interesting to note here that the predicted
{$L_{FIR}$} is generated mainly by the absorption of Lyman continuum and Lyman
$\alpha$ photons, since the extinction affecting the continuum is very
weak. The efficiency in the destruction of ionizing photons has been
included in our models as a constant fraction of 30\%, as discussed
above. This efficiency controls both the resulting {$L_{FIR}$} and {$L(H\beta)$}\
values. The good agreement between predictions and observations allows us
to rely on the efficiency we have assumed, at least for some individual
cases. The good agreement with the observed FIR luminosity implies also
that the burst is dominating the energetics of the whole region. This is
also evident when comparing the predicted and observed optical
continua. The contribution of any underlying older stellar population to
the total optical continuum is negligible when compared to the burst
emission.

\subsection{IZw~18 (=Mrk116)} 

IZw~18 is the prototype of an isolated low metallicity blue compact galaxy
experiencing an intense episode of star formation. With a mean metallicity
as low as Z = {Z$_\odot$} /40 (Searle \& Sargent 1972) it constitutes an ideal
laboratory to study the formation and evolution of massive stars at very
low metal abundances, and has been studied extensively in the last
years. Assuming a single instantaneous burst, Lequeux et al. (1981)
concluded that the present star formation episode in IZw~18 is not older
than 3~Myr, but noted that an extended burst during around 15~Myr would
also be consistent with the observational data. Kunth, Matteucci \& Marconi
(1995) have analyzed the chemical evolution of this metal deficient
galaxy. Their models favour an extended star formation process active at
least during the last 10-20 Myr, instead of a very young instantaneous
episode. Martin  (1996) has found an expanding shell of ionized gas
around the massive stellar cluster. From the kinematics of this shell and
from the analysis of chemical evolution models she concluded that
massive stars have been forming continuously in IZw18 during the last 13
Myr. Finally, Hunter \& Thronson (1995) have identified individual stars in
IZw~18 with WFPC-2 on the Hubble Space Telescope. The picture emerging from
their results is one in which star formation began some tens of Myr ago,
especially in the central and northern part of the galaxy. Star formation
should have continued since then throughout the rest of the galaxy in
smaller star--forming events, giving rise to two main concentrations of
massive stars, comprising around half of them, together with other stars
scattered throughout the galaxy. The age estimated for the inner {H$\alpha$}
filaments span between 1 and around 10~Myr. An additional outer filament
could be several tens of Myr old and is probably related to the onset of
star formation in IZw~18.

Our models can reproduce the majority of the observational parameters
either by assuming an instantaneous burst 3.0 Myr old, in agreement with
Lequeux et al. (1981), or an extended star formation process with a
constant SFR during the last 13 Myr. Considering the
additional constraints imposed by the chemical evolution models and HST
imaging, we have considered only the extended burst scenario. The onset of
relatively small star formation events throughout the galaxy at a nearly
constant rate would indeed mimic a continuous star formation process if
analyzed globally, as we do through the large IUE aperture. The low
metallicity of this object does not allow to measure the individual \ion{Si}{iv}
and \ion{C}{iv} lines of stellar origin, so that we cannot constrain in any way the
IMF slope. We have hence assumed a standard Salpeter IMF in our analysis.

Both the continuum and the emission lines are weakly reddened, with {$E(B-V)$}\
around 0.05 in both cases. Although at such low reddening our sensitivity
to the shape of the extinction law is small, the best fit is obtained
assuming the Galactic law ($\epsilon$ = 0.06-Gal, 0.10-LMC, 0.12-SMC). We wonder
therefore whether this reddening could be due to residual Galactic
extinction, not completely corrected by the Burstein \& Heiles (1984)
method. The predicted FIR emission is very close to the limit of
detectability of IRAS, explaining so its non-detection by this satellite.
The predicted radio spectral index, mainly thermal, coincides well with the
observational one (although the radio emission is underestimated by an
order of magnitude). The synthetic {H$\beta$} emission is also consistent with
the value we measured and is essentially identical to the one derived from
the HST observations by Hunter \& Thronson (1995).

It can be seen in Fig.~\ref{spec1} that the synthetic continuum reproduces
well both the observed UV and optical ranges, up to 7000~\AA. There is no
significant excess continuum in the optical as in other galaxies. We
therefore conclude that the present burst of star formation is dominating
the emission over the whole UV-optical range, and that the continuum
associated to an underlying population if any is negligible.  The
contribution from stars formed in a prior burst several tens of Myr ago, as
postulated by Hunter \& Thronson (1995), should therefore be very small
(compared to the emission from the present burst), even in the optical
range. Our conclusion is that while the {H$\alpha$} filaments show that star
formation could have begun in IZw~18 several tens of Myr ago, the global
star formation rate should have increased significantly around 13~Myr ago,
and has been maintained essentially constant since then.

Our models predict a very small number of WR stars in IZw~18, with {$WR/(WR+O)$}\
$\approx$ 3~10$^{-4}$. Legrand et al. (1997b) and Izotov et al. (1997)
detected a small population of WR stars in this galaxy. While both groups
agree that the optical bumps are due to at least some WC stars, the
 presence of
WN ones is more controversial. Our models do not predict the presence of WC
stars at such low metallicities, unless the formation of WR stars in binary
systems is considered. Cervi\~no (1998) shows that even for a small
fraction of stars evolving in binary systems, mass transfer allows the
formation of WRs (mainly of WC type) at ages above around 6~Myr, in good
agreement with our age estimates. On the other hand, models assuming enhanced
mass loss also predict the formation of a small number of WC stars at very
low metallicities (Schaerer \& Vacca 1998), within a short time period of
1~Myr peaking at an age of 3~Myr. These predictions would be in agreement with
the observations only if these stars have been formed within one of the
very young clusters in IZw~18. A precise location of these stars would be
required to disentangle which effect is dominating in this case: the
formation of WRs via the binary channel, or enhanced mass loss rates (or
both together!). Up to now, there is only a marginal detection with HST of
2 possible WR candidates by Hunter \& Thronson (1995), and some evidences
compatible with 5--9 WNL stars and/or compact nebular \ion{He}{ii} emission by de
Mello et al. (1998).  

Evolutionary models also predict a small number of red supergiant stars
(RSG) at the metallicity of IZw~18, and indeed some individual K and M RSGs
have been identified by Hunter \& Thronson (1995). Cervi\~no \& Mas-Hesse
(1994) explained how current stellar evolutionary tracks predict smaller
populations of RSGs the lower the metal abundance. The relatively blue $V-K$
value measured by Thuan (1983) through a circular aperture 8$\arcsec$ in diameter
($V-K$ = 0.57$\pm$0.23) is well reproduced by our models ($V-K$ $\approx$
0.5$\pm$0.1 for an extended burst between 10 and 20 Myr old), thus
supporting the predictions of current stellar tracks.

\subsection{Haro~2 (=Mrk~33)} 

Haro~2 is one of the brightest blue compact galaxy in our sample. At a
distance of 20.2 ~Mpc it has the shape and brightness profile of an
elliptical galaxy but with a bright, blue nucleus containing an important
number of WR stars (Kunth \& Joubert 1985). At a metallicity 3 times lower
than the Sun it shows clear stellar absorption lines in the UV range (\ion{Si}{iv}
and \ion{C}{iv}). In the optical spectrum strong stellar absorption wings are also
clearly visible around the Balmer emission lines. Recently, Lequeux et
al. (1995) and Legrand et al. (1997a) have found evidences of a strong
outflow of gas having been ejected from the starburst region at velocities
close to 200 km/s, around 1~Myr ago.

Fanelli et al. (1988) have analyzed the UV spectrum of Haro~2 using
optimized, non-evolutionary, stellar synthesis techniques. These authors
derived a strongly discontinuous luminosity function for this galaxy, with
an important contribution to the UV emission of main sequence stars in the
ranges O3-6 (46\% at 1500~\AA ) and B1-2 (48\%), as well as a marginal
contribution from A0-2 stars (6\%).  The interstellar reddening affecting
the UV continuum derived by this technique corresponds to {$E(B-V)$} = 0.15
(assuming a Galactic-like extinction law). The strong contribution of the
O3-6 group reveals the presence of an important population of young massive
stars (less than 5~Myr old). From the discontinuities in the derived
luminosity functions they concluded that the present burst of star
formation has been preceded by at least two bursts, the most recent of
which ended not more than 20 Myr ago giving an important contribution of
B1-B2 stars.  Our analysis, based on evolutionary synthesis techniques,
yields an age of 4.5~Myr for the present burst, with best fits obtained for
a rather steep IMF ($\alpha \geq 3.0$) and a nearly instantaneous
star-formation episode. Our fitting procedure yields an extinction law
similar to the LMC one, with {$E(B-V)$} =~0.12. Both Galactic and SMC laws are
excluded, with $\epsilon$ parameters above 0.14, compared to $\epsilon$ = 0.06 for the
LMC law. The age and extinction strength we derive are consistent with
Fanelli et al. (1988) for the present burst. Nevertheless, our models
naturally predict an important contribution to the UV continuum of stars
with spectral types cooler than O6, with no need to invoke a discontinuous
luminosity function.  The discontinuity in the luminosity function found by
the optimizing synthesis technique could be a consequence of adopting a
Galactic-like extinction law, which we found unable to reproduce the
observations in Haro~2.

On the other hand, our models clearly underestimate the optical continuum,
as shown in Fig.~\ref{spec1}. It seems that the stars formed in the present
burst, the only ones to which our UV observables are sensitive, account for
no more than 60\% of the total optical emission. The shape of the optical
continuum seems to be dominated by A stars which must have been formed in
a previous much older burst.
Our models correctly reproduce several other independent observables of
Haro~2. We predict a relative population of WR stars {$WR/(WR+O)$} = 0.04, which
can account for the observed WR features. The FIR emission is also well
explained supporting the extinction value we have derived from our UV
fitting procedure. The excess FIR emission is probably associated to the
underlying stellar population, which, as we have shown, is significant in
this galaxy. The total {H$\beta$} emission is also satisfactorily reproduced
within a factor 1.3, as well as the relatively low effective
temperature. The {$WR/(WR+O)$} and {$T_{eff}$} values derived for this galaxy set indeed
strong constraints against an extended star formation episode. 

Our results are also consistent with the detection of outflowing gas cited
above. As discussed by Lequeux et al. (1995), the gas should have been
ejected from the central part of the galaxy around 1~Myr ago. This is
consistent with a relatively evolved, massive starburst, which begins to
ignite supernovae after the first 3.5~Myr of evolution (assuming
{$M_{up}$} $\approx$ 120~{M$_\odot$}). Since the mass transformed into stars in
Haro~2 is rather high (more than 6~10$^6$ {M$_\odot$} ), the estimated number of
supernova explosions in the first 10$^6$ yr after the
most massive stars have exhausted their nuclear fuel has to be as high as
500-1000. The kinetic energy ejected by so many supernovae is large enough
to power the outflowing wind. Moreover, such a strong release of energy
could have blown away much of the gas left in the core of this
galaxy, explaining the non-detection by Lequeux et al. (1995) of any
neutral gas at the redshift of the starburst. 

\subsection{Mrk~710 (=NGC~3049)}

Mrk~710 is the only large spiral galaxy with a starburst nucleus in our
sample. It is furthermore one of the two objects with a metallicity as high as
solar, to be compared with the properties of star-formation processes in
metal-deficient blue compact galaxies. Mrk~710 is known to host an
important population of more than 15000 WR stars (Kunth \& Schild 1986,
Vacca \& Conti 1992). As expected for a starburst located within the bulge
of a large galaxy, the relative contribution of the underlying population
is very important. We can see in Fig.~\ref{spec1} that around 60\% of the
observed optical continuum emission is associated to this older
population, whose contribution to the UV below 2000~\AA\ is negligible. 
The measured {$W(H\beta)$} value has therefore to be corrected by a
factor of 2.5, in order to derive the value intrinsically associated to the
present starburst. An additional correction for differential extinction was
also applied, yielding a final {$W(H\beta)$} value almost a factor of 3 higher than
the observed one.

The observational parameters of Mrk~710 can be in general well reproduced
by an instantaneous episode of star formation, around 3.5~Myr old, with an
IMF somewhat flatter than Salpeter one. The predicted WR stars population,
with {$WR/(WR+O)$} = 0.06, reproduces well the observed WR features. The large
{$L(WR)/L(H\beta)$} ratio is indeed consistent with the prediction for solar metallicity
stars. It cannot be reproduced by extended star formation models and
constrains the episode to have been essentially instantaneous. On the
other hand, the effective temperature derived from the optical emission
lines is clearly underestimated by the models. The origin of this
discrepancy could be the presence of some WR stars with effective
temperatures higher than those predicted by current evolutionary
tracks. There are indeed observational evidences that some WR stars can
reach {$T_{eff}$} values above 10$^5$~K. If some of them are present, they could
dominate the shape of the integrated ionizing continuum. On the other hand,
the detection of the OI~$\lambda$6300~\AA\ line in this galaxy indicates
that shocks are affecting the ionization process so that the
{$T_{eff}$} values derived assuming pure photoionization  might be overestimated.
Due to its high metallicity, Mrk~710 shows also the strongest \ion{Si}{iv} 
and \ion{C}{iv}
absorption lines in our sample, supporting the correlation between the
equivalent widths of these lines and the nebular metal abundance we have
plotted in Fig.~\ref{wsc}. 

Despite the close to solar metallicity of this galaxy, the extinction
affecting the UV continuum differs significantly from the Galactic
extinction law. We show in Fig.~\ref{red} the synthetic continuum optimally
reddened by using three possible laws. It can be seen that
reasonable fits are achieved only by using the LMC or SMC laws, with the
best fit obtained with the LMC one for {$E(B-V)$} = 0.17. The Galactic law is
completely excluded.  The FIR luminosity is underestimated by 30\%, as
expected in this galaxy where the IRAS data are integrated over a much
larger region than the IUE flux, and include the whole nucleus and the
spiral arms. Nevertheless, the fact that the observed FIR luminosity is
within a factor 1.3 of the predicted one indicates that the starburst in the
nucleus of this galaxy is the most powerful source of energy dominating
over other possible individual star formation episodes in the spiral arms.

The mass of gas transformed into stars is close to 10$^6$~{M$_\odot$}. While
this amount of mass is clearly higher than in Giant \ion{H}{ii} regions in spiral
arms, it is similar to the values derived in some blue compact galaxies. We
conclude therefore that there are no significant differences between the
bursts observed in relatively small, metal deficient, structureless
galaxies, like BCGs, and the episodes taking place in the nuclei of large
metal-rich spiral galaxies such as Mrk~710.

\subsection{NGC~3256}

NGC~3256 has been classified as the remnant of two colliding galaxies, on
the basis of its tidal tails and highly chaotic nuclear region (Graham et
al. 1984). Joseph \& Wright (1985) concluded that such a merger would
produce a burst of star formation of exceptional intensity, what they
called a ``super starburst''. The detection of strong \ion{Si}{iv} and \ion{C}{iv}
absorption lines by Joseph et al. (1986) is clearly indicative of
the presence of massive stars. Moorwood \& Oliva (1994) have mapped the spatial
distribution of several near infrared lines, finding that Br$\gamma$ is
extended over a 26$\arcsec\times13\arcsec$ 
region. Doyon, Joseph \& Wright (1994) have
detected a large number of red supergiants in the central part of the
galaxy, constraining the age of the starburst to between 12 and 27 Myr,
assuming continuous star formation episodes with different decaying star
formation rates. We included this galaxy in our sample for comparison to
analyze the properties of a very intense episode of star formation in a
dusty medium and induced by a merging process.

The UV continuum and {$W(H\beta)$} value are best reproduced by a rather evolved
cluster around 5~Myr old, in a very dusty environment ({$E(B-V)$} = 0.16
for an SMC law) and with an IMF slope close to Salpeter one. The derived
amount of gas transformed into stars is indeed very large, M $>$ 10$^7$
{M$_\odot$} and the measured extinction is significantly larger than in normal
blue compact galaxies. It is interesting to note that in Mrk~710 and
NGC~3256, the two galaxies with higher metalllicity and the more affected
by reddening in our sample, the extinction seems to follow LMC or SMC laws
characterized by a weak bump at 2175~\AA, supporting our conclusion that
there is no correlation between metallicity and the shape of the extinction
law. 

While our models correctly reproduce the observed UV continuum, they
significantly underestimate the total FIR emission, as well as the total
{H$\beta$} luminosity by almost an order of magnitude. Moreover,
at an age around 5~Myr the cluster would be too young to host the significant
number of RSGs found there. The large extinction suggests that a
significant number of massive stars could be completely hidden by dust, so
that while their contribution to the UV continuum would be negligible,
almost all their emission would be reradiated in the FIR range.  Their
ionizing flux could also contibute to the Br$\gamma$ emission observed in
the near infrared, as well as to the Balmer lines emission. If this is the
case, our analysis would correspond to only a group of massive stars which
happen to be less affected by extinction than the bulk of the cluster, so
that the results can not be extrapolated to the starburst as a whole. The
mass of stars derived by Doyon et al. (1994) from the near infrared lines
is indeed also a factor of 10 larger than the mass we derive from the UV
continuum flux, supporting the idea that only a fraction of the stars are
visible in the UV. This scheme is also favoured by the fact that the
extinction derived from the Balmer decrement is much larger than the value
derived from UV continuum fits ({$E(B-V)$} = 0.85 vs. 0.16, respectively). While
no contribution at all is expected from the hidden stars in the UV, we
would expect some contribution from the obscured regions at {H$\alpha$}
wavelengths, yielding a large global Balmer decrement.

We conclude that starbursts taking place in dusty regions of merging
galaxies can have a rather different appearance from those of BCGs: flat UV
continua due to the strong extinction, reddened emission lines, hidden
massive stars, etc... These are the properties expected for the starburst
powering ultraluminous IRAS galaxies, in several of which the UV continuum
could even not be detected.  Nevertheless, the intrinsic properties of the
star formation processes  are similar to those of
the dust--free weaker bursts observed in other objects.

\section {Conclusions} 

We have analyzed the properties of the star formation episodes taking place
in a sample of blue compact and irregular galaxies. We have applied our
evolutionary population synthesis models to a multiwavelength data set,
including independent UV, optical, FIR and radio parameters related to
different physical processes. Comparing the predicted with the
observational values of the different parameters we have been able to
constrain the age, the star formation regime (instantaneous or extended)
and the Initial Mass Function slope of the massive star formation processes
powering these galaxies, as well as the shape and strength of the
interstellar extinction affecting them. We summarize our main results 
as follows:

\begin{itemize}

\item  Formation of massive stars proceeds by episodic, very short bursts, so
that the stars can be considered to be essentially coeval. Nevertheless,
the superposition of several of these short bursts at different times could
mimic in some cases a star formation process extended during several
million years when the integrated emission from an object is analyzed.

\item  Massive star clusters with nearly instantaneous bursts have ages below
5-6 Myr with an average around 3.5~Myr.  Older bursts would not produce
enough ionizing flux to be detected as emission line galaxies. Powerful
IRAS galaxies might be Postburst Galaxies, characterized by a strong FIR\
emission with very weak emission lines.

\item  Despite the large range in the number of stars formed and the amount of
gas transformed (covering 4 orders of magnitude) the derived IMF shows a
slope in average very close to the solar neighbourhood value. The IMF does not
seem to depend onto the various star formation scenarios in objects as
different as giant \ion{H}{ii} regions, BCGs, starburst nuclei and merging
galaxies. We have found no correlation between the deviations from this
mean value and the metallicity, morphology or interaction state of these
galaxies.

\item  Emission associated to these starbursts completely dominates the UV,
optical and FIR ranges, indicating that the present
bursts might have been one of the strongest in their history or in some
cases the first global episode of massive star formation. In some galaxies
however the contribution of the young stars to the optical continuum is
comparable to the emission associated to older stars formed in preceding
bursts.

\item  No starbursts have been found with {$W(H\beta)$} as high as predicted by the
models for the first 2~Myr after massive star have been formed, despite the
evidences that stars with initial masses close to 100~{M$_\odot$} form in these
galaxies with an IMF slope similar to that of the solar neighbourhood. The
fact that already during accretion stars more massive than 40~{M$_\odot$} might
burn a substantial fraction of its central hydrogen could lead to smaller
effective temperatures and ionizing powers in the first 2~Myr, possibly
explaining so this effect.

\item  Low mass stars formation should take place in these regions before the
most massive stars ignite, and wouldn't be therefore strictly coeval to
them.  Otherwise, they would end their slow collapse when the most massive
stars are already evolved and the ionizing flux of the cluster has almost
completely faded, not being any longer recognized as a starburst.

\item  The observed population of Wolf-Rayet stars is well reproduced by current
models. The detection of large {$WR/(WR+O)$} ratios is the strongest indication of
a short duration star formation episode.

\item  The relatively blue $V-K$ colors measured in IZw~18, which has been forming
massive stars during several Myr, confirms the predictions of current
stellar evolutionary tracks that very few red supergiants would be formed
at low metallicities.

\item  A relatively large number of supernova explosions (up to 1 per 100 years)
is expected to take place in these compact regions. Thousands of supernova
remnants have to be present in the most evolved of these starbursts. The
release of kinetic energy and the ejection of metals must severely disturb
the interstellar gas producing bubbles and metal enriched outflowing winds.

\item  The good agreement between observed and predicted FIR luminosities
excludes in general the possibility that a significant number of massive
stars could be hidden within UV-optical opaque clouds, although this
probably happens in some individual objects. 

\item  The extinction follows a law similar to the LMC and SMC ones, with a very
weak absorption bump at 2175~\AA, if any. There is no dependence between
the shape of the extinction law and the metallicity or the size of the
regions, indicating that the strong radiation associated to the burst has
destroyed the dust components responsible for the UV bump. We find no need to
invoke a specific extinction law other than the LMC or SMC ones.

\item  The strength of the extinction derived from the Balmer decrement is in
many cases larger than the one obtained from the UV continuum. While the
contamination by strong stellar absorption lines, with equivalent widths
around 4-5~\AA\, could explain the discrepancy in some cases, it is more likely
related to the inhomogeneous distribution of stars, ionized gas and
dust. Dust particles concentrate along pat\-chy structures far away from the
continuum sources, but still within the ionized gas, affecting 
more the emission lines than the continuum. The trend toward higher values
of the nebular extinction, the more evolved the regions, suggests that
the newly formed dust grains are ejected far away from the progenitor
stars.

\end{itemize} 

\begin{acknowledgements}
We wish to thank  M. Cervi\~no for his help in the development of
the synthesis models. JMMH is grateful for the hospitality of the Institut
d'Astrophysique de Paris, where part of this work has been performed.  We
are grateful for the support provided by the staff of the VILSPA-IUE,
Nan\c cay and Roque de los Muchachos (La Palma island) observatories.  We
wish to particularly thank the collaboration by L. Bottinelli and
L. Gouguenheim for performing the observations at Nan\c cay. We are
grateful to  M. Rosa and S. D'Odorico who provided to us their spectrum of
NGC~5471. Extensive
use of the NASA Extragalactic Database is also acknowledged.  This work has
been partially financed through different grants from Spanish CICYT
(ESP95-0389-C02-02) and from the Spanish-French PICASSO program.
\end{acknowledgements}

\end{document}